\pdfoutput=1
\documentclass[aps,prd,amsmath,floats,floatfix, twocolumn,
superscriptaddress,nofootinbib,showpacs,longbibliography]{revtex4-1}
\usepackage[T1]{fontenc}
\usepackage[utf8]{inputenc}
\usepackage{txfonts}
\usepackage{verbatim}
\usepackage[dvipsnames, usenames]{xcolor}
\definecolor{linkcolor}{rgb}{0.0,0.3,0.5}
\usepackage[hypertexnames=false, unicode, colorlinks=true, linkcolor=linkcolor,
citecolor=linkcolor, filecolor=linkcolor,urlcolor=linkcolor,
pdfusetitle]{hyperref}
\usepackage[all]{hypcap}
\usepackage{graphicx}
\usepackage{xspace}
\usepackage{amssymb}
\usepackage{microtype}

\usepackage{url}

\usepackage[english]{babel}
\usepackage{blindtext}

\usepackage[normalem]{ulem} 
\usepackage{bm} 
\usepackage{mathrsfs}

\usepackage[caption=false]{subfig}

\DeclareMathAlphabet{\mathpzc}{OT1}{pzc}{m}{it}

\usepackage[nomessages]{fp}

\newcommand{\model}[1]{\texttt{gwNRHME\_NRSur\_q4}}

\newcommand{\orcid}[1]{\href{https://orcid.org/#1}{\includegraphics[width=10pt]{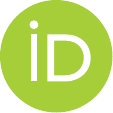}}}
\begin{document}
\title{Including higher-order modes in a quadrupolar eccentric numerical relativity surrogate using universal eccentric modulation functions}

\author{Tousif Islam\,\orcid{0000-0002-3434-0084}}
\email{tousifislam@ucsb.edu}
\affiliation{Kavli Institute for Theoretical Physics, University of California Santa Barbara, Kohn Hall, Lagoon Rd, Santa Barbara, CA 93106, USA}

\author{Adhrit Ravichandran\, \orcid{0000-0002-9589-3168}}
\affiliation{Department of Mathematics, Center for Scientific Computing and Data Science Research, University of Massachusetts, Dartmouth, MA 02747, USA}

\author{Peter James Nee\, \orcid{0000-0002-2362-5420}}
\affiliation{Max Planck Institute for Gravitational Physics (Albert Einstein Institute), Am Mühlenberg 1, 14476 Potsdam, Germany}

\author{Scott E. Field\,\orcid{0000-0002-6037-3277}}
\affiliation{Department of Mathematics, Center for Scientific Computing and Data Science Research, University of Massachusetts, Dartmouth, MA 02747, USA}

\author{Vijay Varma\,\orcid{0000-0002-9994-1761}}
\affiliation{Department of Mathematics, Center for Scientific Computing and Data Science Research, University of Massachusetts, Dartmouth, MA 02747, USA}

\author{Harald P. Pfeiffer\,\orcid{0000-0001-9288-519X}}
\affiliation{Max Planck Institute for Gravitational Physics (Albert Einstein Institute), Am Mühlenberg 1, 14476 Potsdam, Germany}

\author{Andrea Ceja\, \orcid{0000-0002-1681-7299}}
\affiliation{Nicholas and Lee Begovich Center for Gravitational-Wave Physics and Astronomy, California State University Fullerton, Fullerton, California 92834, USA}

\author{Noora Ghadiri}
\affiliation{Nicholas and Lee Begovich Center for Gravitational-Wave Physics and Astronomy, California State University Fullerton, Fullerton, California 92834, USA}
\affiliation{The Grainger College of Engineering, Department of Physics \& Illinois Center for Advanced Studies of the Universe, University of Illinois Urbana-Champaign, Urbana, Illinois 61801, USA}

\author{Lawrence E. Kidder\,\orcid{0000-0001-5392-7342}}
\affiliation{Cornell Center for Astrophysics and Planetary Science, Cornell University, Ithaca, New York 14853, USA}

\author{Prayush Kumar\,\orcid{0000-0001-5523-4603}}
\affiliation{International Centre for Theoretical Sciences, Tata Institute of Fundamental Research, Bangalore 560089, India}

\author{Marlo Morales\,\orcid{0000-0002-0593-4318}}
\affiliation{Nicholas and Lee Begovich Center for Gravitational-Wave Physics and Astronomy, California State University Fullerton, Fullerton, California 92834, USA}

\author{Abhishek Ravishankar\,\orcid{0009-0006-6519-8996}}
\affiliation{Department of Mathematics, Center for Scientific Computing and Data
Science Research, University of Massachusetts, Dartmouth, MA 02747, USA}

\author{Antoni Ramos-Buades\,\orcid{0000-0002-6874-7421}}
\affiliation{Departament de Física, Universitat de les Illes Balears, IAC3 – IEEC, Crta. Valldemossa km 7.5, E-07122 Palma, Spain}

\author{Katie Rink\,\orcid{0000-0002-1494-3494}}
\affiliation{Department of Physics and Weinberg Institute for Theoretical Physics, University of Texas at Austin, TX 78712, USA}

\author{Hannes R. Rüter\,\orcid{0000-0002-3442-5360}}
\affiliation{CENTRA, Departamento de Física, Instituto Superior Técnico,
Universidade de Lisboa, Avenida Rovisco Pais 1, 1049-001 Lisboa, Portugal}

\author{Mark A. Scheel\,\orcid{0000-0001-6656-9134}}
\affiliation{Theoretical Astrophysics 350-17, California Institute of Technology, Pasadena, CA 91125, USA}

\author{Md Arif Shaikh\,\orcid{0000-0003-0826-6164}}
\affiliation{Department of Physics, Vivekananda Satavarshiki Mahavidyalaya
(affiliated to Vidyasagar University), Manikpara 721513, West Bengal, India}

\author{Daniel Tellez}
\affiliation{Nicholas and Lee Begovich Center for Gravitational-Wave Physics and Astronomy, California State University Fullerton, Fullerton, California 92834, USA}

\hypersetup{pdfauthor={Islam et al.}}
\date{\today}

\begin{abstract}
\texttt{gwNRHME} is a framework that converts multi-modal (i.e., containing several spherical harmonic modes) quasi-circular waveforms into their eccentric counterparts, provided the quadrupolar eccentric mode is known, by exploiting universal eccentric modulation functions. Leveraging this framework, we combine the quasi-circular NR surrogate model \texttt{NRHybSur3dq8} with the quadrupolar, non-spinning, eccentric surrogate \texttt{NRSurE\_q4NoSpin\_22} to construct a multi-modal, non-spinning, 
eccentric model, denoted as \model{}, which includes nine modes: 
$(2,\{1,2\})$, $(3,\{1,2,3\})$, $(4,\{2,3,4\})$, and $(5,5)$. When compared against 156 eccentric SXS NR waveforms, \model{} achieves median frequency-domain mismatches (computed using the Advanced LIGO design sensitivity) of $\sim 9\times 10^{-5}$, with a standard deviation of $\sim 2 \times 10^{-4}$.
To demonstrate the modularity of the framework, we further combine \texttt{NRSurE\_q4NoSpin\_22} with effective-one-body (EOB) models \texttt{SEOBNRv5HM} and \texttt{TEOBResumS-Dali} in their non-spinning limits, yielding  eccentric waveforms with median mismatches of $\sim 2\times10^{-4}$ and $\sim 10^{-3}$, respectively, with standard deviation of $\sim 2 \times 10^{-3}$ and $\sim 2 \times 10^{-2}$ respectively. 
Finally, we provide both a surrogate model, \texttt{gwEccEvolve\_q4NoSpin\_Sur}, and an analytical model, \texttt{gwEccEvNSv2}, for the eccentricity evolution up to $2M$ before merger, based on eccentricity definitions derived from the universal modulation 
functions. The \texttt{gwNRHME} framework is publicly available through the \texttt{gwModels} package, and the resulting waveform models will be released via the \texttt{gwsurrogate} package.

\end{abstract}
\maketitle

\section{Introduction}
Detecting gravitational waves (GWs) from eccentric binary black holes (BBHs) remains a key objective in GW astronomy. This pursuit is motivated by two main reasons. First, most BBH mergers detected so far are well described by quasi-circular templates~\cite{Harry:2010zz,LIGOScientific:2014pky,VIRGO:2014yos,KAGRA:2020tym,LIGOScientific:2018mvr,LIGOScientific:2020ibl,LIGOScientific:2021usb,LIGOScientific:2021djp,LIGOScientific:2025slb,LIGOScientific:2026qni}. Although some events show potential signatures of eccentricity~\cite{Romero-Shaw:2020thy,Gayathri:2020coq,Gamba:2021gap,Ramos-Buades:2023yhy,Gupte:2024jfe,Morras:2025xfu,romeroshaw2025gw200208222617eccentricblackholebinary,Planas:2025jny,Tiwari:2025fua,Kacanja:2025kpr,Jan:2025fps,Phukon:2025cky,McMillin:2025hof}, a clear consensus has not been reached owing to the lack of accurate eccentric waveform models that incorporate essential physical effects such as spin precession and higher-order (beyond-quadrupole) modes. Second, astrophysical population synthesis studies suggest that most eccentric binaries circularize before their GW emission enters the sensitivity band ($\sim 15$–$20\,\mathrm{Hz}$) of current detectors. However, a subset of systems formed in dense environments, such as globular clusters or galactic nuclei, through dynamical captures or hierarchical triple interactions may retain measurable eccentricity near merger~\cite{Rodriguez:2017pec,Rodriguez:2018pss,Samsing:2017xmd,Zevin:2018kzq,Zevin:2021rtf,Samsing:2020tda}. Observing such systems would provide unique insights into their dynamical formation channels and the environments in which they reside, as their GW signatures are expected to encode information about these astrophysical conditions.

To ensure that eccentric binaries are not missed during detection and can be accurately characterized once observed, it is crucial to develop reliable models that includes eccentricity. In recent years, considerable progress has been made toward this goal, with hundreds of numerical relativity (NR) simulations performed~\cite{Scheel:2025jct,Mroue:2010re, Healy:2017zqj,Buonanno:2006ui,Husa:2007rh,Ramos-Buades:2018azo,Ramos-Buades:2019uvh,Purrer:2012wy,Bonino:2024xrv,Ramos-Buades:2022lgf,Nee:2025zdy,Healy:2022wdn} to enhance our understanding of eccentric BBH dynamics and to aid in the construction of accurate models. 
These numerical efforts have been complemented by theoretical studies investigating the role of eccentricity in BBH dynamics and radiation properties within the post-Newtonian (PN) approximation (cf.\ Ref.~\cite{Blanchet:2013haa} for a review). Together, they have led to the development of semi-analytical eccentric waveform models, including those based on the effective-one-body (EOB) formalism and phenomenological approaches~\cite{Tiwari:2019jtz, Huerta:2014eca, Moore:2016qxz, Damour:2004bz, Konigsdorffer:2006zt, Memmesheimer:2004cv, Cho:2021oai,Hinderer:2017jcs,Cao:2017ndf,Chiaramello:2020ehz,Albanesi:2023bgi,Albanesi:2022xge,Riemenschneider:2021ppj,Chiaramello:2020ehz,Ramos-Buades:2021adz,Liu:2023ldr,Huerta:2016rwp,Huerta:2017kez,Joshi:2022ocr,Wang:2023ueg,Carullo:2023kvj,Nagar:2021gss,Tanay:2016zog,Gamboa:2024hli,Morras:2025nbp,Morras:2025nlp,Hinder:2017sxy,Planas:2025feq,Chattaraj:2022tay,Setyawati:2021gom,Paul:2024ujx,Manna:2024ycx,Maurya:2025shc,Nagar:2024oyk,Ramos-Buades:2026kbq,Morras:2026fho,Thomas:2026qrg}. Another line of work employs PN-guided, data-driven decomposition techniques to extract subharmonics induced by eccentricity within the EOB framework and to construct computationally efficient waveform models~\cite{Patterson:2024vbo,Islam:2025rjl,Islam:2025bhf,Islam:2025llx}. While these models incorporate the effects of eccentricity during the inspiral phase (the early stage of the BBH dynamics), nearly all of them assume that the binary has circularized by the time of merger. In addition, these models do not incorporate any calibration to eccentric NR simulations.

A complementary direction was explored in Ref.~\cite{Islam:2021mha}, which introduced a fully data-driven surrogate model, \texttt{NRSur2dq1Ecc}, constructed from NR simulations and restricted to equal-mass eccentric BBH systems. Recently, a new framework has been proposed~\cite{Nee:2025nmh}, incorporating several modelling improvements as well as generalising to unequal mass binaries, resulting in the surrogate model \texttt{NRSurE\_q4NoSpin\_22}. These models rely minimally on analytical assumptions and do not impose circularity at merger, thereby capturing the full nonlinear dynamics encoded in the NR simulations, unlike their semi-analytical counterparts. On the other hand, by utilizing a large set of eccentric BBH NR simulations from various catalogs, both in the spinning and non-spinning limits, Refs.~\cite{Islam:2024rhm,Islam:2024bza} have empirically identified simple, model-independent universal relations among different spherical harmonic modes in the waveform. These relations were subsequently used to construct a general framework, \texttt{gwNR(X)HME}, which enables the conversion of a quasi-circular, multi-modal waveform model into an eccentric, multi-modal waveform model, provided that a quadrupolar eccentric model is available~\cite{Islam:2024rhm,Islam:2024bza}. As we focus on the non-spinning limit in this work, we drop the \texttt{X} and refer to the framework simply as \texttt{gwNRHME}. As a proof of principle, Ref.~\cite{Islam:2024zqo} employed this framework to develop a hybrid, non-spinning eccentric model by combining the multi-modal, quasi-circular NR surrogate model \texttt{NRHybSur3dq8}~\cite{Varma:2018mmi} with the quadrupolar, non-spinning PN eccentric model \texttt{EccentricIMR}~\cite{Hinder:2017sxy}. These universal relations among different modes also provide a robust and smoothly varying measure of eccentricity that can be extracted directly from the gravitational waveforms~\cite{Islam:2025oiv}, offering an alternative to other eccentricity estimators~\cite{Moore:2016qxz,Gopakumar:1997bs,Damour:2004bz,Konigsdorffer:2006zt,Tessmer:2008tq,Arun:2009mc,Tanay:2016zog,Shaikh:2023ypz,Shaikh:2025tae}.

In this paper, we employ the \texttt{gwNRHME} framework to construct an accurate, multi-modal, non-spinning eccentric model, \texttt{gwNRHME\_NRSur}, by combining the state-of-the-art quadrupolar NR surrogate \texttt{NRSurE\_q4NoSpin\_22} with the multi-modal, quasi-circular NR surrogate model \texttt{NRHybSur3dq8} (in its non-spinning limit). The resulting model, \model{}, by construction, makes no approximations related to circularity at merger and exactly recovers the underlying \texttt{NRSurE\_q4NoSpin\_22} behavior in the quadrupolar limit. For the higher-order modes, the accuracy of \model{} is limited only by the accuracy of the empirical relations established in Refs.~\cite{Islam:2024rhm,Islam:2024bza}. 

The remainder of this paper is organized as follows. In Section~\ref{sec:framework}, we describe the framework and methodology underlying our model construction. Section~\ref{sec:results} presents a detailed assessment of the accuracy and performance of \model{}.  
Next, in Section~\ref{sec:eccevolve_models}, we build a surrogate model for the eccentricity evolution close to merger, following the framework developed in Ref.~\cite{Islam:2025oiv}. We denote this new model as \texttt{gwEccEvolveSur\_NoSpinq4}.
Section~\ref{sec:other_models} then explores the modularity of \texttt{gwNRHME} by applying it to introduce eccentric multi-modal corrections into other state-of-the-art quasi-circular models constructed within the EOB framework yielding \texttt{gwNRHME\_SEOB\_q4} and \texttt{gwNRHME\_TEOB\_q4} models. 
Finally, in Section~\ref{sec:conclusion}, we summarize our findings, discuss the current limitations of our models, and outline possible directions for future improvements. We make our frameworks and models publicly available through \texttt{gwModels}~\footnote{\href{https://github.com/tousifislam/gwModels}{https://github.com/tousifislam/gwModels}} package.

\begin{figure}
\includegraphics[width=\columnwidth]{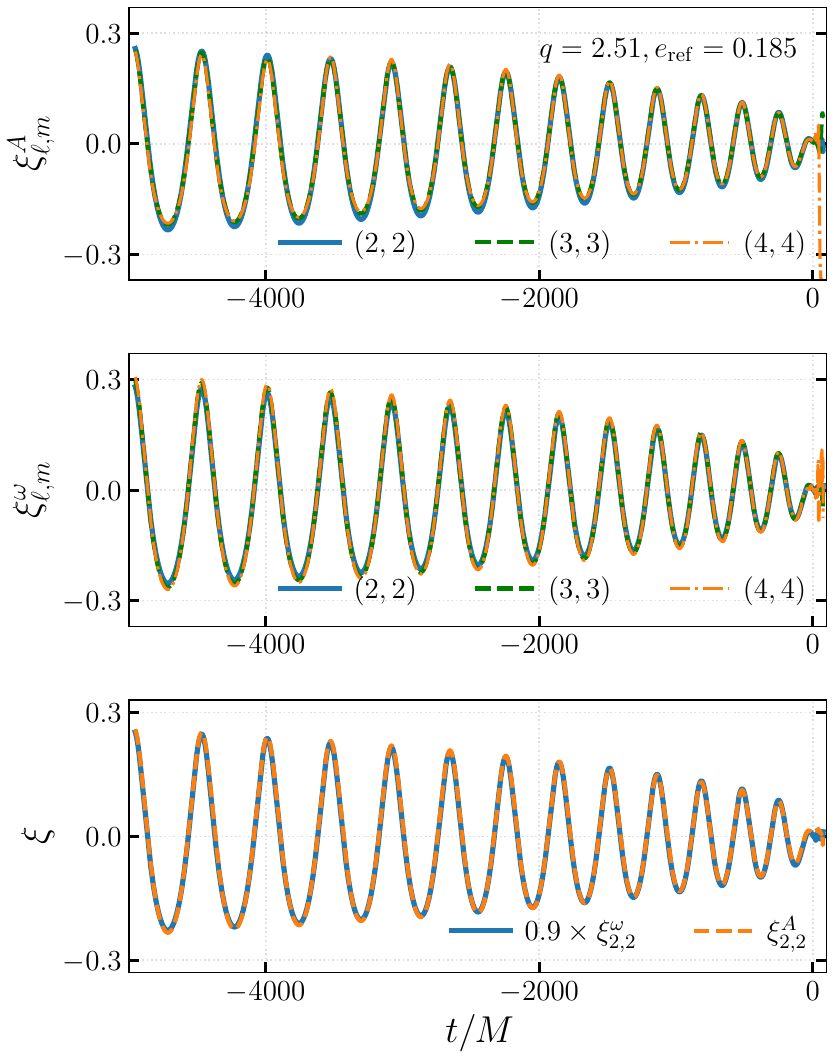}
\caption{We show the eccentric modulation in the amplitudes, $\xi_{\ell m}^{A}$ (upper panel, defined in Eq.(\ref{eq:amp_mod})), and in the instantaneous frequencies, $\xi_{\ell m}^{\omega}$ (middle panel, defined in Eq.(\ref{eq:freq_mod})), for three representative spherical harmonic modes: $(2,2)$ (blue), $(3,3)$ (green), and $(4,4)$ (orange). These modulations are extracted from a NR simulation of a binary with mass ratio $q = 2.51$ and reference eccentricity $e_{\rm ref} = 0.185$, measured at the beginning of the waveform. The lower panel illustrates that the two modulation measures are simply related by a proportionality constant $B = 0.9$ shown in Eq.(\ref{eq:amp_to_freq}). Further details are provided in Section~\ref{sec:universal_modulation}.}
\label{fig:case_115_modulations}
\end{figure}

\section{Modelling framework}
\label{sec:framework}
We begin by decomposing the complexified gravitational waveform $h(t)$ into a superposition of spin-weighted spherical harmonic modes with spin weight $s=-2$ and indices $(\ell, m)$:
\begin{align}
h(t, \theta, \phi; \boldsymbol{\lambda}) 
  &= \sum_{\ell=2}^{\infty} \sum_{m=-\ell}^{\ell} 
     h_{\ell m}(t; \boldsymbol{\lambda}) \, {}_{-2}Y_{\ell m}(\theta,\phi).
\label{hmodes}
\end{align}
Here, $t$ denotes the time coordinate, while $\theta$ and $\phi$ specify the angular position on the sky relative to the merger. The set of intrinsic parameters, $\boldsymbol{\lambda}$, characterizes the binary system and includes quantities such as the component masses and spins. For non-spinning eccentric BBHs, the parameter set is defined as $\boldsymbol{\lambda} := \{q, e_{\rm ref}, l_{\rm ref}\}$, where $q$ is the mass ratio, and $e_{\rm ref}$ and $l_{\rm ref}$ denote the reference eccentricity and mean anomaly, respectively. Each spherical harmonic mode is further expressed in terms of a real-valued amplitude $A_{\ell m}(t)$ and phase $\phi_{\ell m}(t)$ as: $h_{\ell m}(t; q, e_{\rm ref}, l_{\rm ref}) = A_{\ell m}(t) \, e^{i \phi_{\ell m}(t)}$. The instantaneous angular frequency is then defined as: $\omega_{\ell m}(t) = \frac{d\phi_{\ell m}(t)}{dt}$. We set the time corresponding to the maximum amplitude of the $(2,2)$ mode to be $t=0$ for all models considered in this paper.

\begin{figure}
\includegraphics[width=\columnwidth]{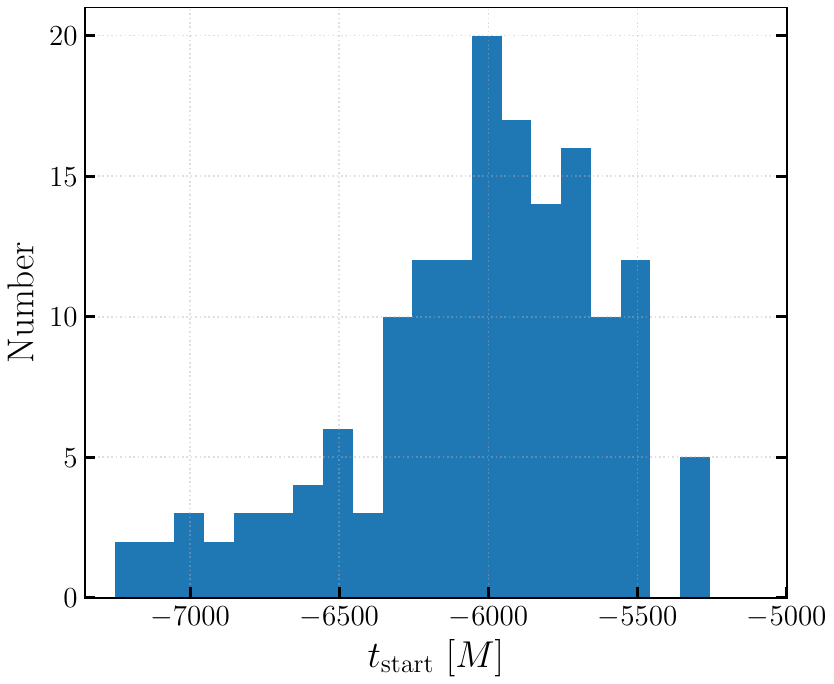}
\caption{We show the distribution of waveform starting times obtained from the quadrupolar eccentric model \texttt{NRSurE\_q4NoSpin\_22}, evaluated at its training data points. Further details are provided in Section~\ref{sec:gwnrhme_framework}.}
\label{fig:22sur_length}
\end{figure}

\subsection{The universal modulation function}
\label{sec:universal_modulation}
Refs.~\cite{Islam:2024rhm,Islam:2024bza} demonstrated that, for non-precessing eccentric BBHs, all spherical harmonic modes exhibit characteristic, mode-independent modulations induced by orbital eccentricity. These modulations can be quantified using either the mode amplitudes, $A_{\ell m}(t; \boldsymbol{\lambda})$, or the instantaneous angular frequencies, $\omega_{\ell m}(t; \boldsymbol{\lambda})$, by comparing them to the corresponding quantities from a quasi-circular waveform characterized by $\boldsymbol{\lambda}^0 := \{q, e_{\rm ref}=0, l_{\rm ref}=0\}$:
\begin{align}
\xi_{\ell m}^{\omega}(t; \boldsymbol{\lambda}) &= b_{\ell m}^\omega 
\frac{\omega_{\ell m}(t; \boldsymbol{\lambda}) - \omega_{\ell m}(t; \boldsymbol{\lambda^0})}
{\omega_{\ell m}(t; \boldsymbol{\lambda^0})}, 
\label{eq:freq_mod} \\
\xi_{\ell m}^{A}(t; \boldsymbol{\lambda}) &= b_{\ell m}^A 
\frac{2}{\ell} \,
\frac{A_{\ell m}(t; \boldsymbol{\lambda}) - A_{\ell m}(t; \boldsymbol{\lambda^0})}
{A_{\ell m}(t; \boldsymbol{\lambda^0})}.
\label{eq:amp_mod}
\end{align}
Here, $b_{\ell m}^\omega$ and $b^{A}_{\ell m}$ are scaling constants chosen to be unity.
Furthermore, Refs.~\cite{Islam:2024rhm,Islam:2024bza} empirically found that the amplitude and frequency modulations are simply related:
\begin{equation}
\xi_{\ell m}^{A}(t; \boldsymbol{\lambda}) \approx B \, \xi_{\ell m}^{\omega}(t; \boldsymbol{\lambda}),
\label{eq:amp_to_freq}
\end{equation}
with a proportionality constant $B \approx 0.9$. In general, $B(t; \boldsymbol{\lambda})$ is a time-dependent function of the intrinsic binary parameters. However, empirical evidence indicates that it can be accurately approximated as a constant~\cite{Islam:2024rhm,Islam:2024bza}.
In Figure~\ref{fig:case_115_modulations}, we show the eccentric modulation functions extracted from a NR simulation of a binary with mass ratio $q = 2.51$ and reference eccentricity $e_{\rm ref} = 0.185$, measured at the start of the waveform. The upper panel shows that the amplitude modulations obtained from different spherical harmonic modes are consistent with one another. The middle panel demonstrates that the frequency modulations are likewise identical across modes. Finally, the lower panel illustrates that the amplitude and frequency modulations are related by a constant scaling factor $B$, demonstrating the universality of the modulation function.
Assuming this relation holds, one can define a common eccentric modulation function. Among different modes, the $(2,2)$ mode yields the most robust and noise-resistant estimate of the modulation function, and we therefore adopt $\xi_{22}^{A}(t; \boldsymbol{\lambda})$ as the common modulation parameter $\xi(t) := \xi_{22}^{A}(t; \boldsymbol{\lambda})$.

\begin{figure}
\includegraphics[width=\columnwidth]{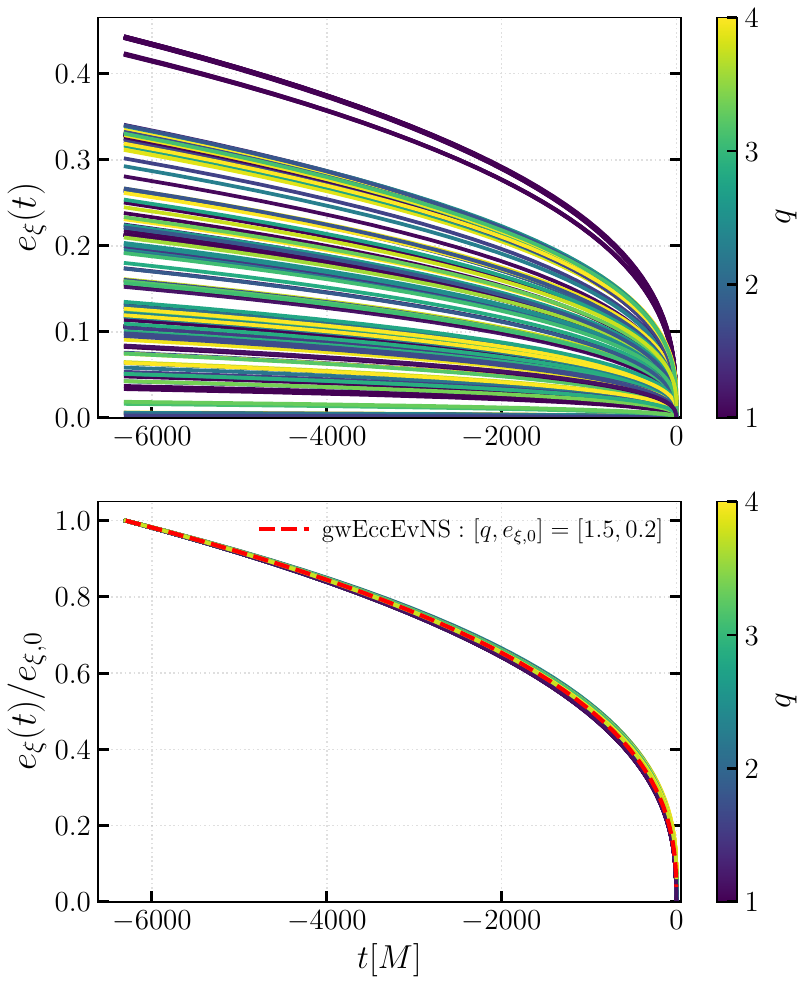}
\caption{(\textit{\textbf{Upper panel:}}) We show the evolution of the eccentricity, $e_{\xi}(t)$ (defined in Eq.(\ref{eq:exi})), estimated for the NR simulations used in this work by applying the framework introduced in Ref.~\cite{Islam:2025oiv} and implemented in the \texttt{gwModels} package (\href{https://github.com/tousifislam/gwModels}{https://github.com/tousifislam/gwModels}). (\textit{\textbf{Lower panel:}}) We scale the eccentricity by its initial value and observe that the eccentricity evolution exhibits a degree of universality across the configurations considered. For comparison, we also include the eccentricity evolution of a binary with parameters $[q, e_{\rm ref}] = [1.5, 0.2]$ (red dashed line), obtained using the \texttt{gwEccEvNS} model presented in Ref.~\cite{Islam:2025oiv}. For both panels, we color–code the lines according to their mass-ratio values.
Further details are provided in Section~\ref{sec:eccmeasure}.}
\label{fig:ecc_xi_evolution}
\end{figure}

\begin{figure}
\includegraphics[width=\columnwidth]{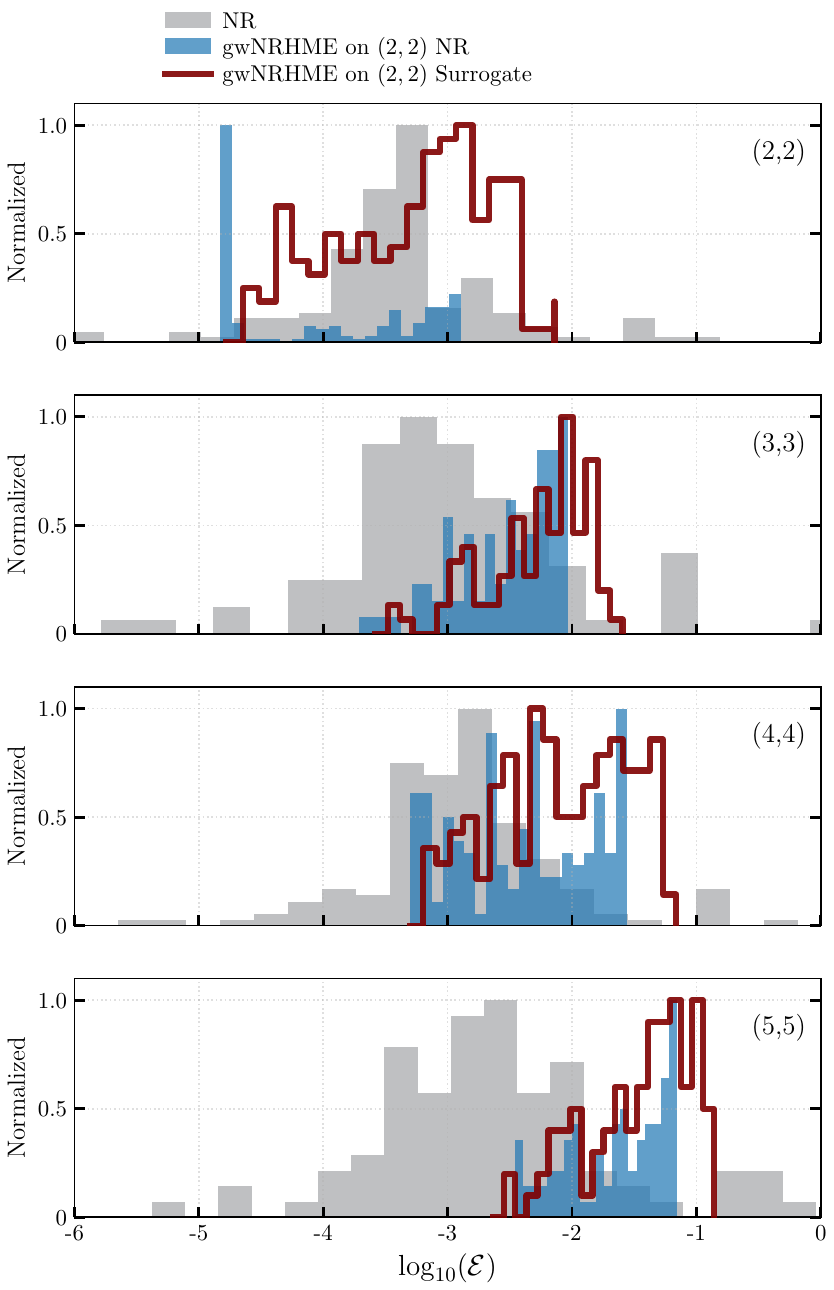}
\caption{We show the distribution of the relative $L_2$-norm errors (defined in Eq.~\ref{eq:l2err}) for spherical harmonic modes with $\ell=m$, comparing the \model{} predictions against $156$ NR waveforms. The grey histogram represents the $L_2$-norm errors between the two highest-resolution NR datasets, serving as a benchmark. The maroon histogram indicates the \model{} errors relative to NR, while the blue histogram shows the errors for higher-order modes obtained by applying the \texttt{gwNRHME} framework to the quasi-circular \texttt{NRHybSur3dq8} model using the universal modulation extracted from the $(2,2)$ NR mode. Further details are provided in Section~\ref{sec:tderrors}.}
\label{fig:l2err_ell_m}
\end{figure}

\begin{figure}
\includegraphics[width=\columnwidth]{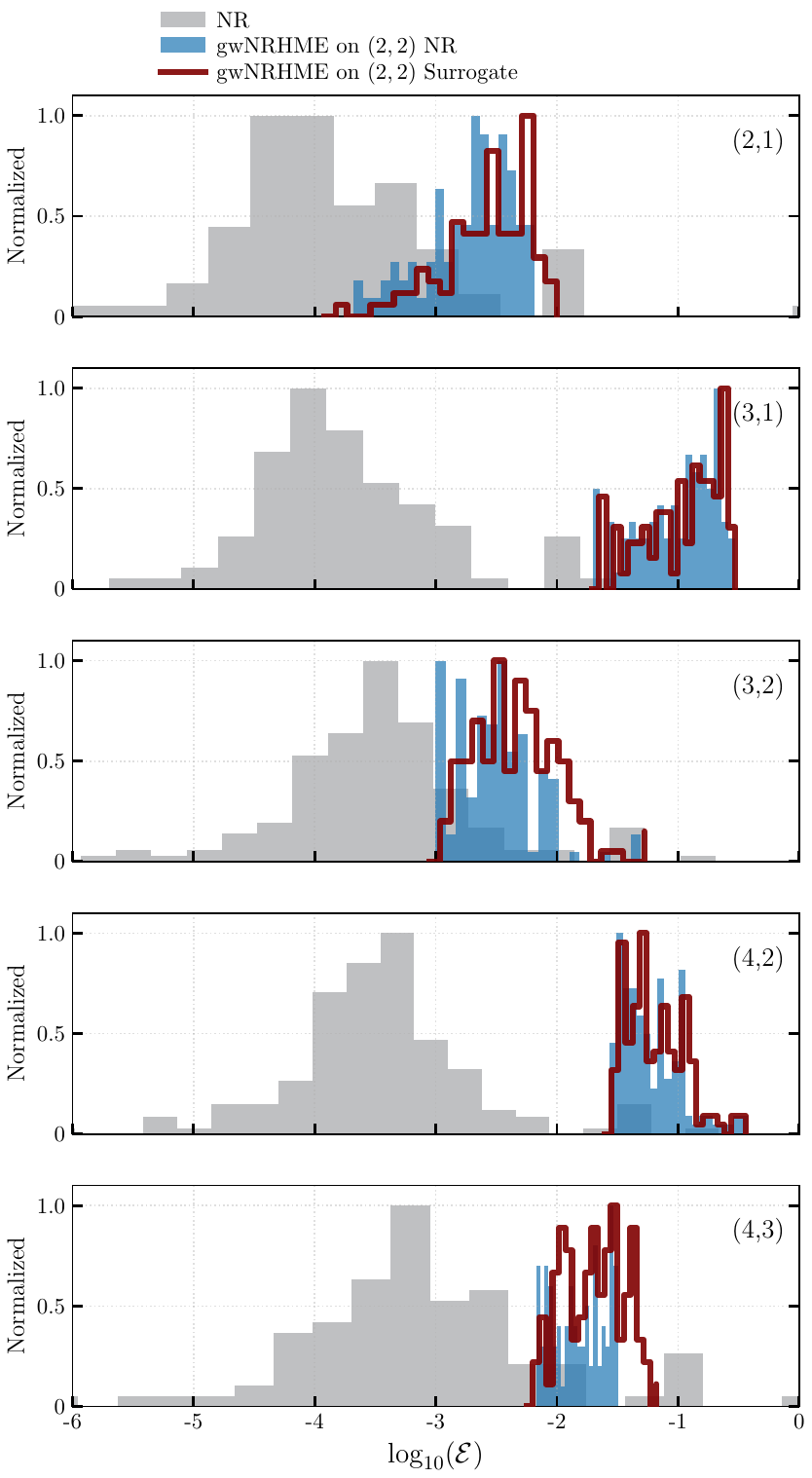}
\caption{Same as Figure~\ref{fig:l2err_ell_m}, but for spherical harmonic modes with $\ell \neq m$. Additional details can be found in Section~\ref{sec:tderrors}.}
\label{fig:l2err_othermodes}
\end{figure}

\subsection{\texttt{gwNRHME} framework}
\label{sec:gwnrhme_framework}
We exploit the universality and consistency of the eccentric modulation functions to construct multi-modal, non-spinning eccentric waveforms employing \texttt{gwNRHME} framework. Specifically, we combine the dominant quadrupolar mode of a non-spinning eccentric NR waveform, $h_{22}(t;\boldsymbol{\lambda})$, with the corresponding quasi-circular, non-precessing, multi-modal NR waveform, $h_{\ell m}(t;\boldsymbol{\lambda}^0)$, to generate higher-order eccentric modes. The amplitudes of the higher-order modes are modeled as
\begin{equation}
A_{\ell m}^{\texttt{gwNRHME}}(t; \boldsymbol{\lambda}) 
= A_{\ell m}(t; \boldsymbol{\lambda}^0) 
\left[ 1 + \frac{\ell}{2} \, \xi(t) \right].
\end{equation}
The corresponding instantaneous frequencies are given by
\begin{equation}
\omega_{\ell m}^{\texttt{gwNRHME}}(t; \boldsymbol{\lambda}) 
= \omega_{\ell m}(t; \boldsymbol{\lambda}^0) 
\left[ 1 + \frac{\xi(t)}{B} \right].
\end{equation}
Integrating the frequency yields the phase evolution for each mode:
\begin{equation}
\phi_{\ell m}^{\texttt{gwNRHME}}(t; \boldsymbol{\lambda}) 
= \phi_{0} + \int \omega_{\ell m}^{\texttt{gwNRHME}}(t; \boldsymbol{\lambda}) \, dt,
\end{equation}
where $\phi_{0} = \phi_{\ell m}(t; \boldsymbol{\lambda}^0)$ is the integration constant. Finally, the complex strain for each spherical harmonic mode is reconstructed as
\begin{equation}
h_{\ell m}^{\texttt{gwNRHME}}(t; \boldsymbol{\lambda}) 
= A_{\ell m}^{\texttt{gwNRHME}}(t; \boldsymbol{\lambda}) 
\, e^{i \phi_{\ell m}^{\texttt{gwNRHME}}(t; \boldsymbol{\lambda})}.
\end{equation}
This framework has already been demonstrated using non-precessing eccentric NR data and waveform models presented in Refs.~\cite{Islam:2024rhm,Islam:2024bza,Islam:2024zqo}.

As mentioned earlier, in this work our underlying `carrier' quadrupolar eccentric model is \texttt{NRSurE\_q4NoSpin\_22}~\cite{Nee:2025nmh}, a non-spinning eccentric NR surrogate constructed using Gaussian Process Regression (GPR). This model is developed using $\sim156$ NR simulations employing a radial-phase reparameterization of the BBH dynamics. It captures approximately $42$ GW cycles of the $(2,2)$ mode, with initial times ranging from $-7200M$ to $-5300M$ (Figure~\ref{fig:22sur_length}). The median starting time for these simulations are $-5969.6M$.
The model is trained over the parameter space $q \in [1,4]$ and eccentricities $\in [0.001, 0.25]$ (computed at $t=-3000M$). For the quasi-circular base model, we adopt \texttt{NRHybSur3dq8}~\cite{Varma:2018mmi}, an aligned-spin NR surrogate trained on 104 NR simulations with mass ratios $q \leq 8$ and spin magnitudes $|\chi_{1,2}| \leq 0.8$. It includes all spherical harmonic modes up to $\ell \leq 4$ as well as the $(5,5)$ mode, but excludes the $(4,1)$ and $(4,0)$ modes. Nevertheless, the \texttt{gwNRHME} framework is modular, and the quasi-circular base model can be replaced with any other waveform model of choice.

To compute the common modulation function $\xi(t)$, we evaluate \texttt{NRSurE\_q4NoSpin\_22} to obtain the quadrupolar eccentric waveform and \texttt{NRHybSur3dq8} to generate the corresponding quadrupolar quasi-circular waveform. We ensure that $t=0$ denotes the peak of the $(2,2)$ mode for both the waveforms. Once $\xi(t)$ is determined, we employ the \texttt{gwNRHME} framework to map the quasi-circular higher-order modes of \texttt{NRHybSur3dq8} into their eccentric counterparts. The final eccentric, multi-modal surrogate waveform \model{} includes the following spherical harmonic modes:
$(\ell, m) = \{(2,2), (2,1), (3,1), (3,2), (3,3), (4,2), (4,3), (4,4), (5,5)\}$. Unless otherwise noted, this model is adopted as the default in this work.
Schematically, we write it as:
\begin{align}
\texttt{NRHybSur3dq8} 
&+ \texttt{NRSurE\_q4NoSpin\_22} 
+ \texttt{gwNRHME} \nonumber \\
&\longrightarrow \texttt{gwNRHME\_NRSur\_q4}.
\end{align}

However, as one might expect, replacing the base quasi-circular multi-modal model in this construction leads to a slightly different multi-modal eccentric waveform model. In particular, \texttt{NRHybSur3dq8} can be substituted with state-of-the-art quasi-circular models developed within other modeling frameworks. Examples include perturbation-theory-based models such as \texttt{BHPTNRSur1dq1e4}~\cite{Islam:2022laz} and EOB models such as \texttt{SEOBNRv5HM}~\cite{Pompili:2023tna} and \texttt{TEOBResumS-Dali}~\cite{Nagar:2021gss,Nagar:2024dzj}. 
In the latter part of this paper (Section~\ref{sec:other_models}), we perform this replacement explicitly for \texttt{SEOBNRv5HM} and \texttt{TEOBResumS-Dali}, and present the resulting waveform comparisons. We do not include \texttt{BHPTNRSur1dq1e4} at this stage, as it is trained for mass ratios $q \ge 2.5$, thereby covering only a limited portion of the \texttt{NRSurE\_q4NoSpin\_22} parameter space.

\subsection{Measure of eccentricity}
\label{sec:eccmeasure}
Following Ref.~\cite{Islam:2025oiv}, we further employ the universal eccentric modulation time series $\xi(t; \boldsymbol{\lambda})$ to construct a PN-guided, robust, and smoothly varying eccentricity measure, denoted as $e_{\xi}(t)$. This is achieved by first obtaining continuous representations of the upper and lower envelopes of the common modulation function, denoted as $\xi^{\rm env}_{\rm p}(t)$ and $\xi^{\rm env}_{\rm a}(t)$ respectively, and then taking their average. An appropriate prefactor of $b=2/3$ is included to ensure that $e_{\xi}(t)$ recovers the standard Newtonian eccentricity in the low-eccentricity limit. We express our eccentricity measure as~\cite{Islam:2025oiv}:
\begin{equation}
e_{\xi}(t) = b \frac{\xi^{\rm env}_{\rm p}(t) + \xi^{\rm env}_{\rm a}(t)}{2}.
\label{eq:exi}
\end{equation}
Details of this eccentricity measure and comparison with similar measures are given in Ref.~\cite{Islam:2025oiv}.
The continuous representations of the envelopes are derived using PN-inspired fits (which includes terms up to $3.5$ PN) that enforce monotonicity. This eccentricity measure has been validated against eccentric waveforms from NR, EOB, and PN frameworks for both spinning and non-spinning binaries. A key advantage of $e_{\xi}(t)$ is that it provides a meaningful estimate of the binary’s eccentricity even in the strong-field regime, close to the merger. We utilize this framework through \texttt{gwModels}~\footnote{\href{https://github.com/tousifislam/gwModels}{https://github.com/tousifislam/gwModels}} package.

In Figure~\ref{fig:ecc_xi_evolution}, we present the eccentricity evolution $e_{\xi}(t)$, estimated for the NR simulations employed in this work, mapped onto a common time grid chosen to span the entire duration of the shortest simulation in the training set. We find that the eccentricities can reach values as large as $0.4$ at $t = -6200M$. Following the procedure outlined in Ref.~\cite{Islam:2025oiv}, we then rescale the eccentricities by their initial values and observe that the resulting evolution exhibits an almost universal behavior, with only a small spread due to variations in the mass ratio. For a quick comparison, we also overlay the eccentricity evolution predicted by the \texttt{gwEccEvNS}~\cite{Islam:2025oiv} model for a binary with $q = 1.5$ and $e_{\rm ref} = 0.2$ and find that it qualitatively matches the data quite well.

For the remainder of this paper, we use the eccentricity $e_{\xi}(t)$ and denote its initial value as the reference eccentricity $e_{\rm ref}$. Under this convention, the model eccentricities vary within the range $e_{\rm ref} \in [0.002, 0.43]$.

\section{Extending quadrupolar non-spinning eccentric model to higher-order modes}
\label{sec:results}
To evaluate the accuracy of the \model{} waveforms, we compute both the time/phase-optimized time-domain relative $L_2$-norm error and the frequency-domain mismatch between the NR data and the \model{} predictions. The relative $L_2$-norm between two waveforms, $h_1(t)$ and $h_2(t)$, is defined as~\cite{blackman2017numerical}
\begin{equation}\label{eq:l2err}
\mathcal{E} = \int_{t_{\rm min}}^{t_{\rm max}} \frac{|h_{1}(t) - h_{2}(t)|^2}{|h_{1}(t)|^2} \, dt,
\end{equation}
where $t_{\rm min}$ and $t_{\rm max}$ denote the start and end times of the waveforms.  
The frequency-domain mismatch $\mathcal{M}$ between two waveforms is given by~\cite{Cutler:1994ys}
\begin{equation}
\mathcal{M} = 1 - \langle h_1, h_2 \rangle,
\end{equation}
where the inner product is defined as
\begin{equation}
\langle h_1, h_2 \rangle = 4\,\mathcal{R} 
\int_{f_{\mathrm{min}}}^{f_{\mathrm{max}}}
\frac{\tilde{h}_1(f)\,\tilde{h}_2^*(f)}{S_n(f)} \, df.
\label{Eq:freq_domain_Mismatch}
\end{equation}
Here, $\tilde{h}(f)$ denotes the Fourier transform of the strain $h(t)$, $^*$ indicates complex conjugation, $\mathcal{R}$ denotes the real part, and $S_n(f)$ is the one-sided power spectral density (PSD). We adopt the design-sensitivity PSD of the Advanced LIGO detector~\cite{KAGRA:2013rdx}, using a frequency range of $f_{\mathrm{min}} = 20~\mathrm{Hz}$ to $f_{\mathrm{max}} = 999~\mathrm{Hz}$.

Note that, when evaluating the model accuracy, we always use the \model{} predictions obtained from the validation step, wherein the corresponding NR data used for testing are excluded from the training set. This ensures that the resulting $L_2$-norm errors and mismatches represent the worst-case performance of \model{}.

\begin{figure}
\includegraphics[width=\columnwidth]{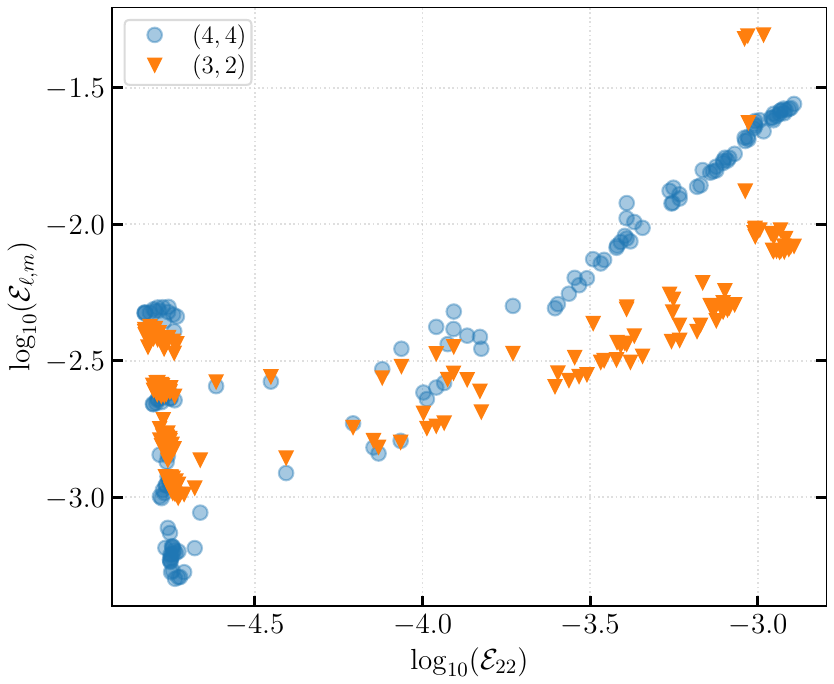}
\caption{We show the correlation between the relative $L_2$-norm errors of the dominant $(2,2)$ mode and two higher-order modes: $(3,2)$ (blue circles) and $(4,4)$ (orange triangles). We find that the errors in the higher-order modes increase whenever the error in the $(2,2)$ mode is large, indicating a clear correlation between the modeling accuracy of the dominant and subdominant modes within \texttt{gwNRHME} framework. Further details are provided in Section~\ref{sec:tderrors}.}
\label{fig:22to44_err_relation}

\end{figure}

\begin{figure*}
\centering
\subfloat[]{
    \includegraphics[width=0.48\textwidth]{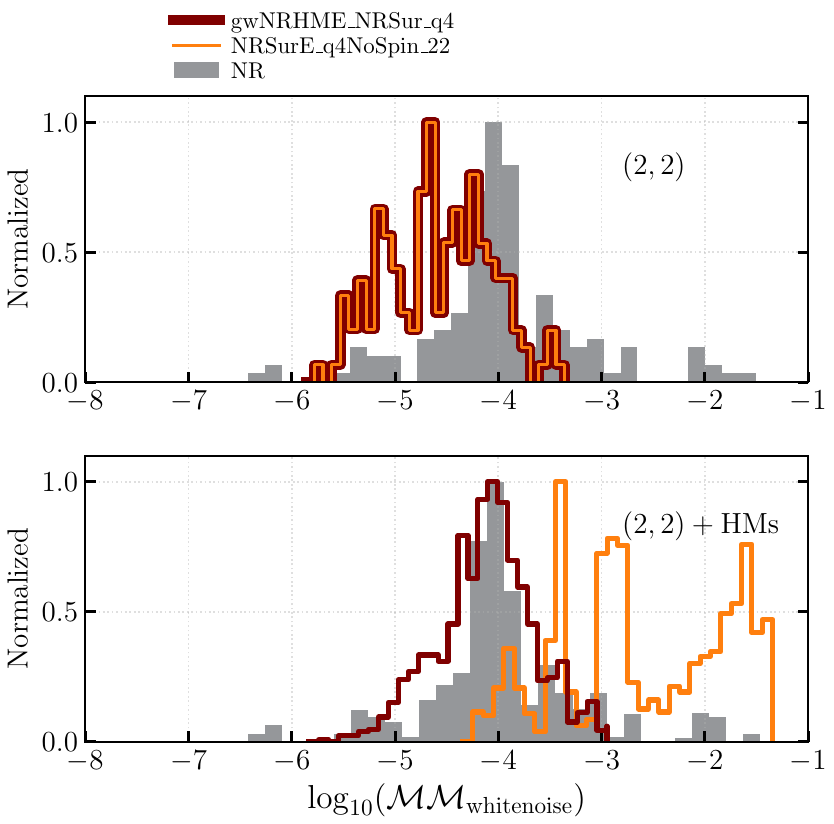}
    \label{fig:mismatches_white}
}
\subfloat[]{
    \includegraphics[width=0.48\textwidth]{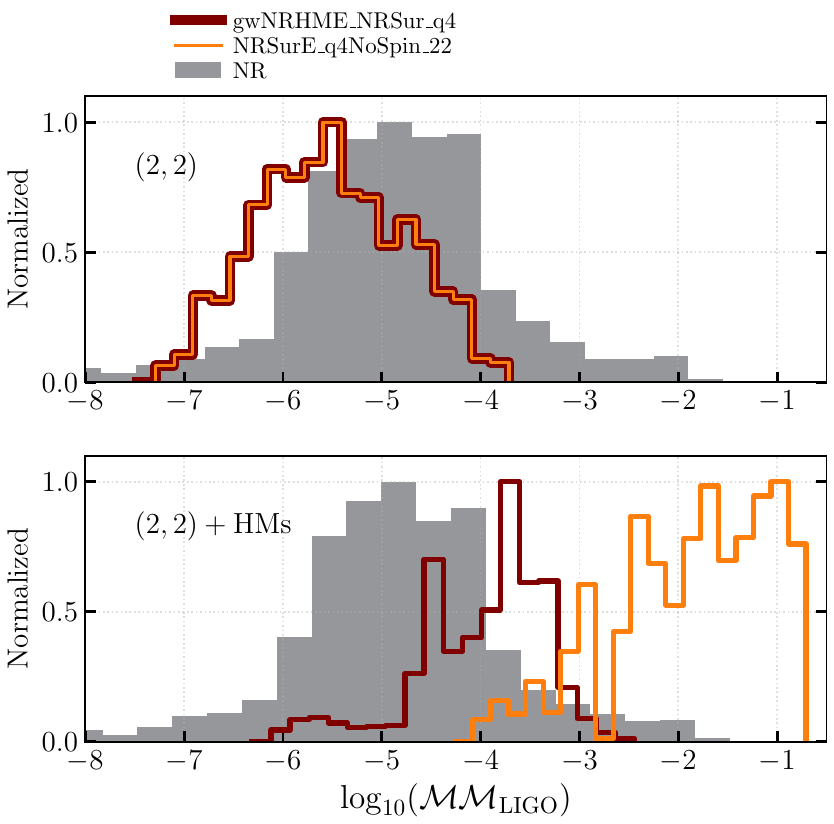}
    \label{fig:mismatches_ligo}
}
\caption{We show the frequency-domain mismatches between the \model{} predictions and NR data (maroon solid histogram), computed assuming both a white-noise spectrum \textbf{(left panels; (a))} and the Advanced LIGO design sensitivity curve \textbf{(right panels; (b))} for different inclinations and orbital phases. Results are shown with and without the inclusion of higher-order modes. For the latter, we include the modes mentioned in Section~\ref{sec:framework}. For comparison, we also present mismatches obtained from NR simulations at two different resolutions (grey histogram) as well as those from the quadrupolar \texttt{NRSurE\_q4NoSpin\_22} model alone. More details are in Section~\ref{sec:freqmismatch}.}
\label{fig:mismatches}
\end{figure*}

\begin{figure*}
\includegraphics[width=\textwidth]{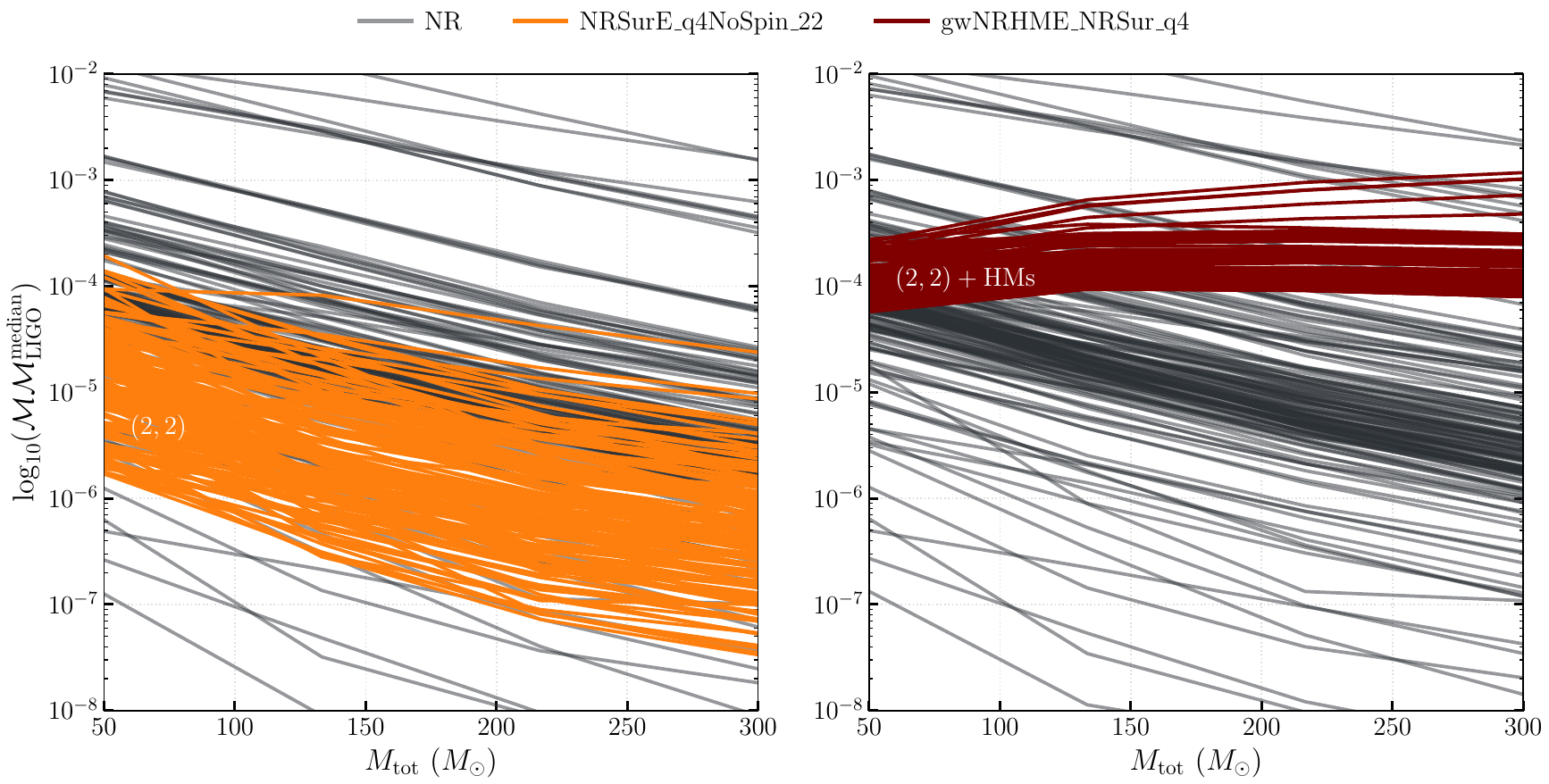}
\caption{We show the median frequency-domain mismatches between surrogate predictions and NR data, computed using the Advanced LIGO design sensitivity curve, as a function of the total detector-frame mass of the BBHs. The \textbf{left panel} shows the median mismatches between the quadrupolar \texttt{NRSurE\_q4NoSpin\_22} model and quadrupolar NR data (orange), while the \textbf{right panel} shows the median mismatches between the \model{} predictions and NR data (maroon), where higher-order modes are included. For reference, we also show the median mismatches between NR waveforms at two different numerical resolutions (grey) without (left panel) and with (right panel) higher-order modes. Further details are in Sec.~\ref{sec:freqmismatch}.}
\label{fig:mismatches_ligo_vs_Mtot}
\end{figure*}

\subsection{Time-domain errors}
\label{sec:tderrors}
In Figures~\ref{fig:l2err_ell_m} and ~\ref{fig:l2err_othermodes}, we present the distribution of the relative $L_2$-norm errors for 9 different spherical harmonic modes, comparing the \model{} predictions against $143$ NR waveforms. 
While Figure~\ref{fig:l2err_ell_m} presents the $L_2$-norm errors for modes with $\ell = m$, Figure~\ref{fig:l2err_othermodes} shows the corresponding results for modes with $\ell \neq m$.
As a benchmark, we also show the $L_2$-norm errors obtained by comparing the two highest-resolution NR simulations. For additional comparison, we include results for higher-order modes constructed by applying the \texttt{gwNRHME} framework to the quasi-circular \texttt{NRHybSur3dq8} model, using the universal modulation extracted from the $(2,2)$ NR mode instead of the \texttt{NRSurE\_q4NoSpin\_22} model.
For the dominant $(2,2)$ mode, the $L_2$-norm errors from \model{} are similar to the NR resolution errors validating the quadrupolar limit of our approach. For higher-order modes, particularly those with $\ell=m$, the errors increase slightly relative to the $(2,2)$ mode but still remain comparable to the NR resolution level. The typical relative $L_2$-norm errors are $\sim10^{-4}$ for the $(2,2)$ mode, $\sim2\times10^{-3}$ for the $(3,3)$ and $(2,1)$ modes, and $\sim10^{-2}$ for the remaining modes. Finally, for the less dominant modes such as $(3,1)$ and $(4,2)$, the errors increase further, ranging from approximately $0.02$ to $0.2$.
We also note that the NR resolution errors exhibit a modest increase for higher-order modes compared to the dominant $(2,2)$ mode.

Next, in Figure~\ref{fig:22to44_err_relation}, we demonstrate a clear correlation between the modeling accuracy of the dominant and subdominant modes, specifically for the $(3,2)$ and $(4,4)$ modes. We find that the errors in these higher-order modes increase whenever the error in the $(2,2)$ mode is large. 
This behavior could arise primarily due to two reasons. 
First, since the eccentricity corrections for the higher-order modes are derived exclusively from the modulation function $\xi(t)$ computed using the amplitude of the $(2,2)$ mode, any inaccuracies present in the $(2,2)$ carrier mode propagate directly into the higher-order modes. 
Second, some of the higher-order modes may exhibit more complex physical effects, such as mode mixing, or involve higher-order corrections that are not currently captured within the \texttt{gwNRHME} framework. 
As these sources of error do not presently limit the overall accuracy of our model, we leave improvements to the treatment of these modes to future work.

\begin{figure}
\includegraphics[width=\columnwidth]{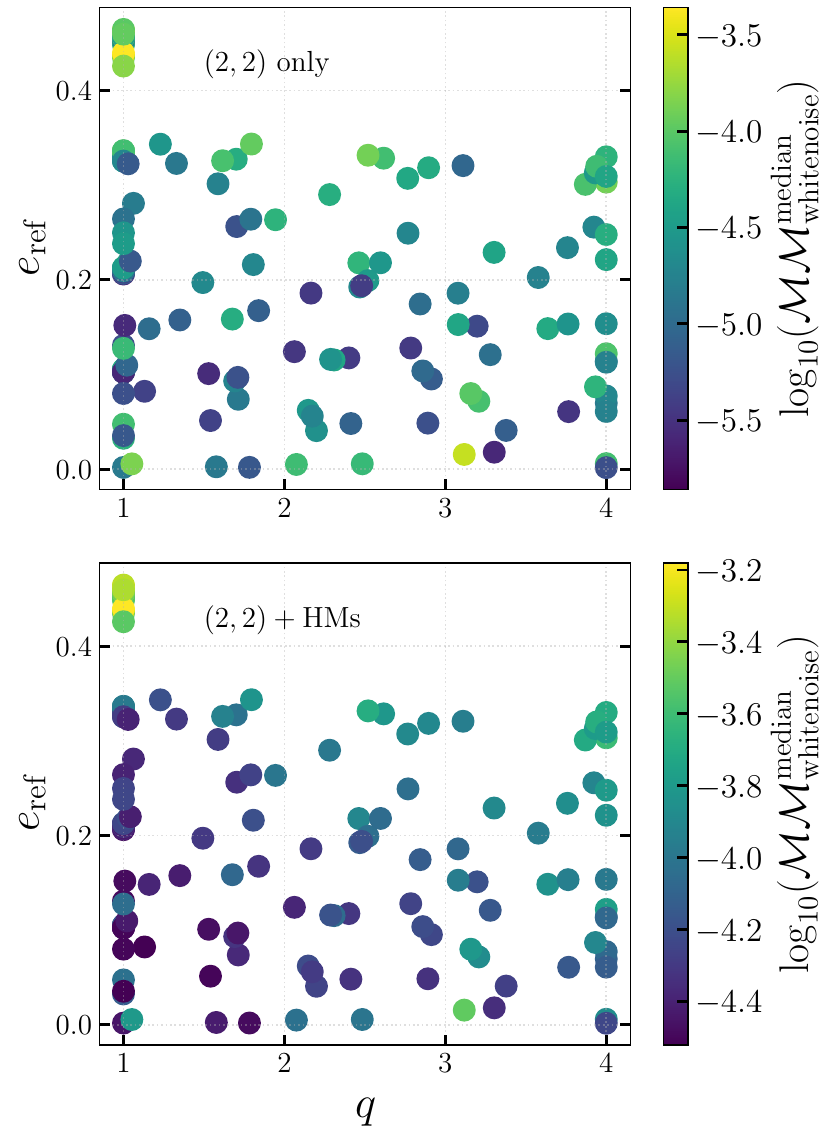}
\caption{We show median white-noise mismatches between the \model{} predictions and NR data as a function of the mass ratio and the reference eccentricity. The median is computed over different values of the orbital phase and inclination. The upper panel shows mismatches computed using only the quadrupolar mode in both \model{} and NR, while the lower panel includes higher-order modes. Further details are provided in Sec.~\ref{sec:results}.}
\label{fig:mismatches_paramspace_colorbar}
\end{figure}

\begin{figure}
\includegraphics[width=\columnwidth]{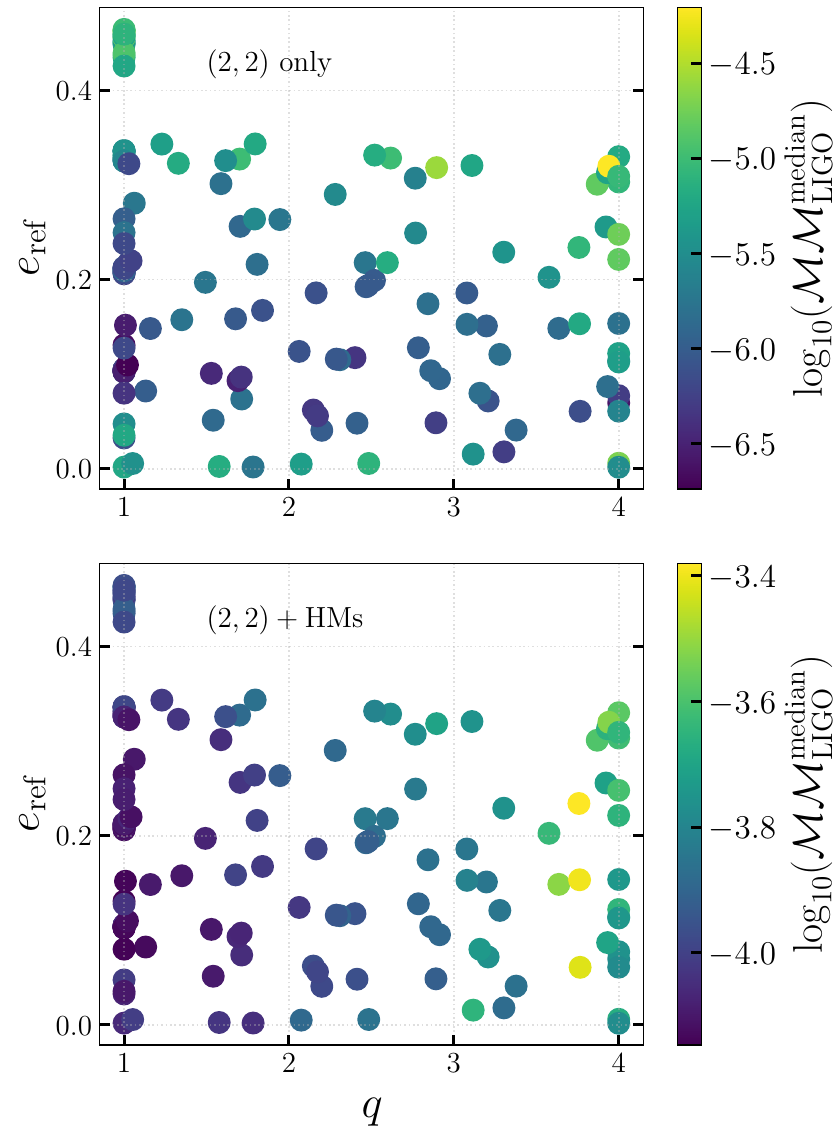}
\caption{We show median Advanced LIGO mismatches between the \model{} predictions and NR data as a function of the mass ratio and the reference eccentricity. The median is computed over different values of the orbital phase and inclination, as well as five detector-frame masses between $50\,M_{\odot}$ and $300\,M_{\odot}$. The upper panel shows mismatches computed using only the quadrupolar mode in both \model{} and NR, while the lower panel includes higher-order modes. Further details are provided in Sec.~\ref{sec:results}.}
\label{fig:mismatches_ligo_paramspace_colorbar}
\end{figure}

\begin{figure}
\includegraphics[width=\columnwidth]{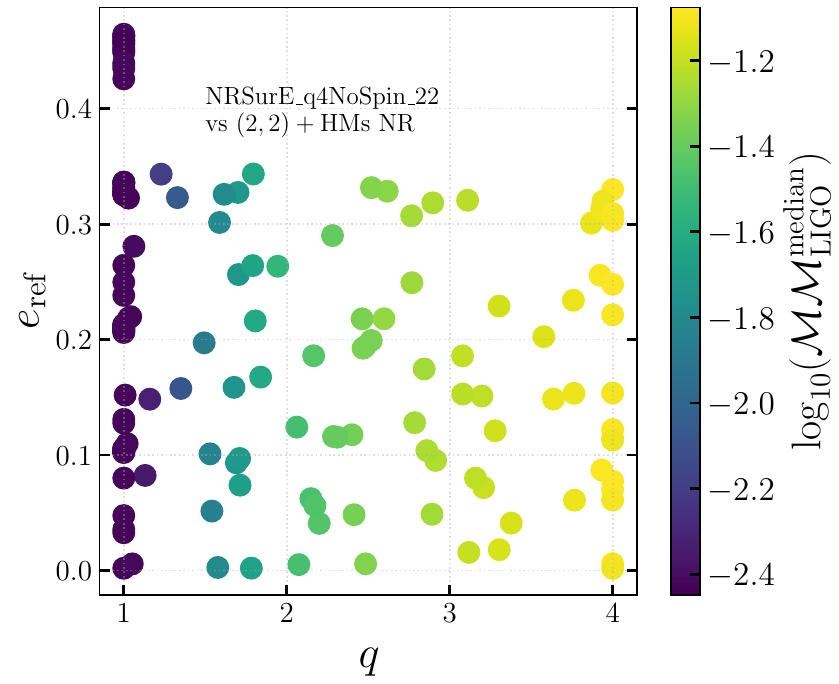}
\caption{Median Advanced LIGO mismatches between the quadrupolar \texttt{NRSurE\_q4NoSpin\_22} predictions and NR data, which include higher-order modes, are shown as a function of the mass ratio and the reference eccentricity. As before, the median is computed over detector-frame masses between $50\,M_{\odot}$ and $300\,M_{\odot}$, as well as over different orbital phases and inclination angles. Unlike in Figs.~\ref{fig:mismatches_paramspace_colorbar} and \ref{fig:mismatches_ligo_paramspace_colorbar}, here the \texttt{NRSurE\_q4NoSpin\_22} waveforms include only the quadrupolar mode, whereas the NR data include higher-order modes. The increased mismatches observed relative to Figs.~\ref{fig:mismatches_paramspace_colorbar} and \ref{fig:mismatches_ligo_paramspace_colorbar} demonstrate the importance of higher-order modes and the necessity of incorporating them into the model. Further details are in Sec.~\ref{sec:results}.}
\label{fig:mismatches_ligo_22sur_vs_NRHM_paramspace_colorbar}
\end{figure}

\begin{figure*}
\includegraphics[width=0.98\textwidth]{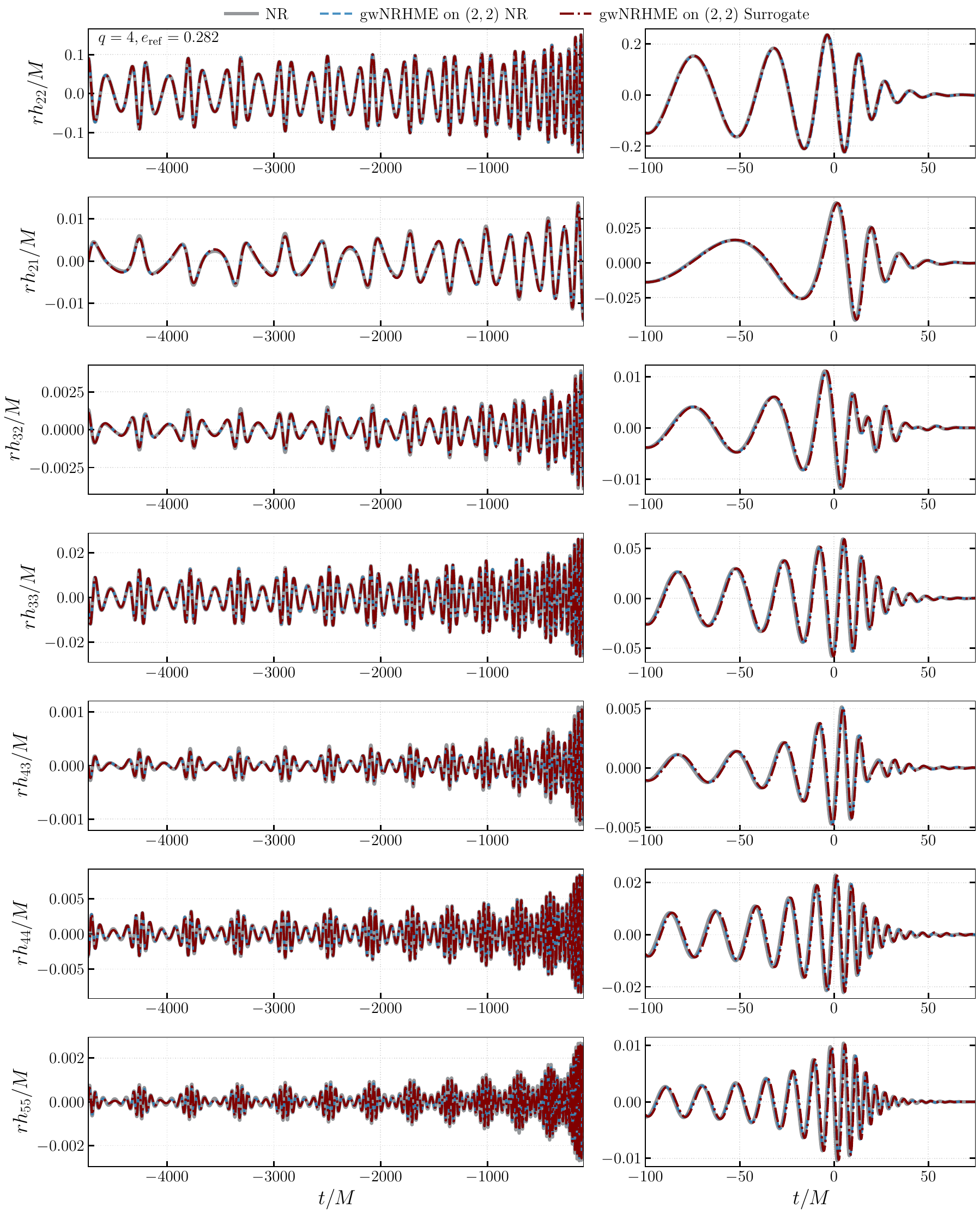}
\caption{We show the eccentric spherical harmonic modes obtained from the \model{} (maroon line) and the corresponding NR data (grey line) for the case that exhibits the largest $L_2$-norm error among most of the higher-order modes (from Figs.~\ref{fig:l2err_ell_m},~\ref{fig:l2err_othermodes}). The \model{} predictions are generated by combining the quadrupolar eccentric model \texttt{NRSurE\_q4NoSpin\_22} and the quasi-circular waveform model \texttt{NRHybSur3dq8} using the \texttt{gwNRHME} framework. For comparison, we also 
show the modes obtained by applying the \texttt{gwNRHME} framework to the quasi-circular \texttt{NRHybSur3dq8} model using the universal modulation extracted from the $(2,2)$ NR mode (blue dashed line). This particular simulation corresponds to a binary with mass ratio $q = 4$ and reference eccentricity $e_{\rm ref} = 0.282$. We find that the \model{} predictions are in agreement with the NR results. More details are in Section~\ref{sec:waveform_comparison}.}
\label{fig:case_0105}
\end{figure*}

\begin{figure*}
\includegraphics[width=0.95\textwidth]{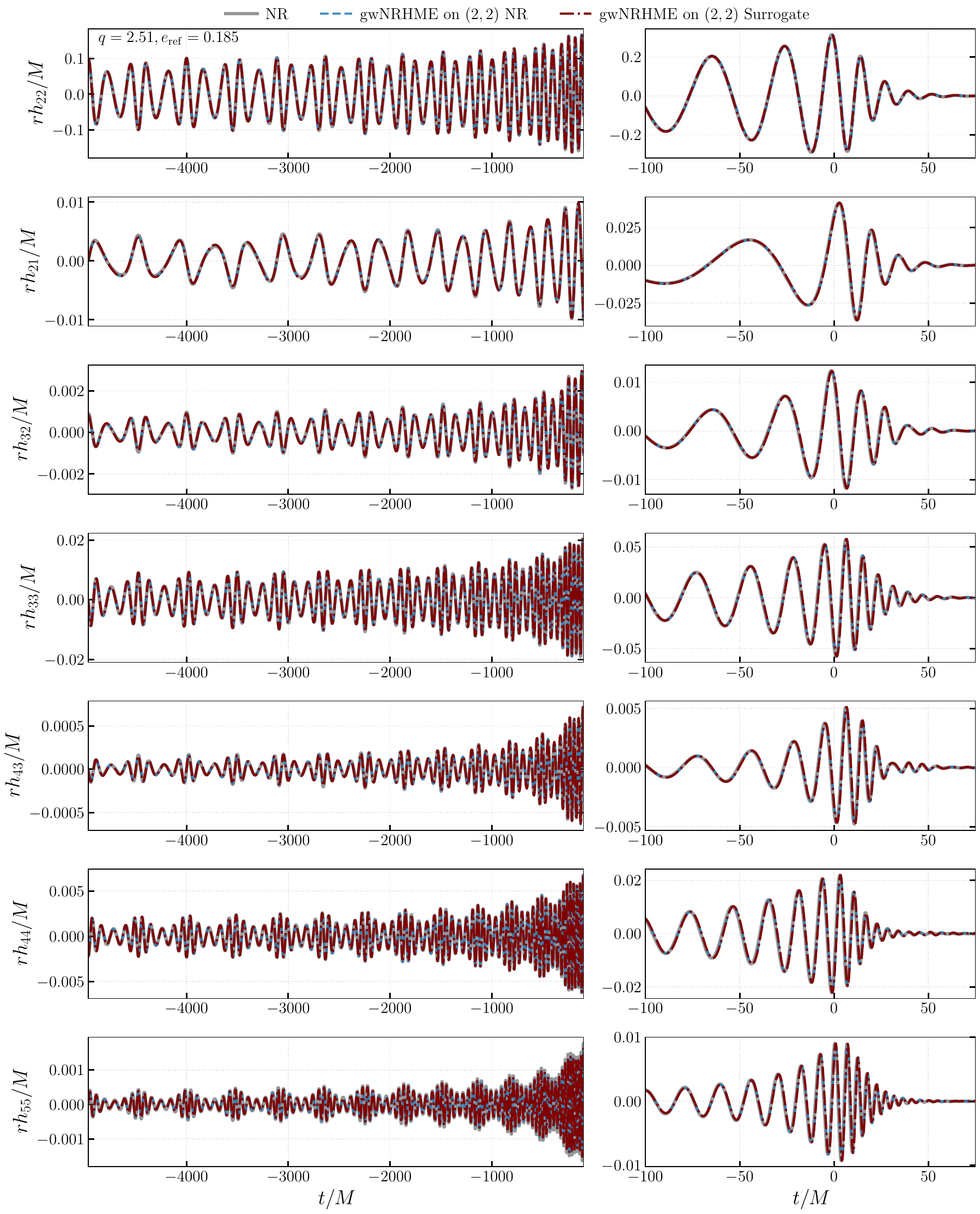}
\caption{We show the eccentric spherical harmonic modes obtained from the \model{} (maroon line) and the corresponding NR data (grey line) for the case that exhibits a typical $L_2$-norm error among most of the higher-order modes. The \model{} predictions are generated by combining the quadrupolar eccentric model \texttt{NRSurE\_q4NoSpin\_22} and the quasi-circular waveform model \texttt{NRHybSur3dq8} using the \texttt{gwNRHME} framework. For comparison, we also 
show the modes obtained by applying the \texttt{gwNRHME} framework to the quasi-circular \texttt{NRHybSur3dq8} model using the universal modulation extracted from the $(2,2)$ NR mode (blue dashed line). This simulation corresponds to a binary with mass ratio $q = 2.51$ and reference eccentricity $e_{\rm ref} = 0.185$. We find that the \model{} predictions are in agreement with the NR results. More details are in Section~\ref{sec:waveform_comparison}.}
\label{fig:case_0115}
\end{figure*}

\subsection{Frequency-domain mismatches}
\label{sec:freqmismatch}
We now analyze the frequency-domain mismatches computed for $17$ different orbital phase and inclination angles. Note that we use the same $17$ orbital phase and inclination angles for both white-noise mismatches as well as for the Advanced LIGO mismatches.
In Figures~\ref{fig:mismatches_white}, we show the mismatches between the \model{} predictions and the NR data (maroon solid histogram), computed assuming a white-noise spectrum, both with and without the inclusion of higher-order modes. For the latter, we include the modes mentioned in Section~\ref{sec:framework}. For comparison, we also present mismatches obtained from NR simulations at two different resolutions (grey histogram), as well as those from the quadrupolar \texttt{NRSurE\_q4NoSpin\_22} model (orange). First, we confirm that the $(2,2)$-mode mismatches from \model{} match exactly with those from the quadrupolar \texttt{NRSurE\_q4NoSpin\_22} model, verifying that \model{} correctly reproduces the latter in the quadrupolar limit. We find that, both with and without higher-order modes, the \model{} mismatches closely overlap with the NR mismatches. The typical mismatches for the $(2,2)$ mode and for the full multi-modal case are $\sim 10^{-5}$. 

To investigate whether the higher-order modes incorporated through the \texttt{gwNRHME} framework in \model{} contribute significantly to the model accuracy, we also compute mismatches between the quadrupolar \texttt{NRSurE\_q4NoSpin\_22} model and the multi-modal NR data. We find that the mismatches for \texttt{NRSurE\_q4NoSpin\_22} against the multi-modal NR data range from $\sim 10^{-3}$ to a few times $10^{-2}$ (orange histogram; lower panel of Figures~\ref{fig:mismatches_white}), indicating only reasonable to poor accuracy. These mismatch values are nearly two orders of magnitude larger than those obtained with \model{}. This clearly demonstrates that explicitly modeling the higher-order modes is essential, and that \model{} is capable of accurately capturing them through the \texttt{gwNRHME} framework.

Next, we compute the Advanced LIGO mismatches for four different detector-frame total masses in the range $M_{\rm tot} \in [50M_{\odot}, 200M_{\odot}]$, considering $17$ different orbital phase and inclination angles for each. In Figure~\ref{fig:mismatches_ligo}, we present the Advanced LIGO mismatches between the \model{} predictions and the NR data, both with and without the inclusion of higher-order modes. 

We find that the \model{} LIGO mismatches lie in the range from $10^{-6}$ to a few times $10^{-3}$ and overlap with the NR mismatches when higher-order modes are included. However, there remains some room for improvements as some of the NR mismatches are even smaller than $10^{-6}$. As before, the Advanced LIGO mismatches between the quadrupolar \texttt{NRSurE\_q4NoSpin\_22} model and NR data (including higher modes) are two to three orders of magnitude larger than those obtained with \model{}.

We then investigate how the mismatches vary as a function of the detector-frame total mass of the binaries. In Figure~\ref{fig:mismatches_ligo_vs_Mtot}, we show the median Advanced LIGO mismatches, computed over different orbital phase and inclination angles, for each of the $156$ NR simulations, as a function of the total mass, both with and without the inclusion of higher-order modes. We find that for \model{}, the median mismatches are typically $\sim 10^{-4}$ for most cases. While the median mismatches for \model{} are slightly larger than those obtained using only the $(2,2)$ mode, they are still sufficiently small for GW data analysis with current ground-based GW detectors.

Finally, to examine how the model accuracy varies across the parameter space, we plot the median mismatches between the \model{} predictions and NR data, computed across different orbital phases, inclination angles, and detector-frame masses, as a function of the mass ratio and reference eccentricity both without and with higher-order modes (see Figures~\ref{fig:mismatches_paramspace_colorbar} and~\ref{fig:mismatches_ligo_paramspace_colorbar}). We find that both the white-noise and Advanced LIGO mismatches increase slightly with increasing mass ratio and eccentricity. This trend is likely due to the denser sampling of training data in the near-equal-mass regime. The largest mismatches are observed near the high–mass-ratio boundary. It is also worth noting that the parameter values corresponding to the largest white-noise and Advanced LIGO mismatches do not necessarily coincide.

We also show the median Advanced LIGO mismatches between the quadrupolar \texttt{NRSurE\_q4NoSpin\_22} predictions and the NR data (including higher-order modes) as a function of the parameter space in Figure~\ref{fig:mismatches_ligo_22sur_vs_NRHM_paramspace_colorbar}. This figure demonstrates that the quadrupolar model \texttt{NRSurE\_q4NoSpin\_22} provides a good approximation to the full NR data only for systems with $q \lesssim 2$. Beyond this mass-ratio threshold, the Advanced LIGO mismatches exceed $10^{-2}$, a commonly adopted benchmark for determining whether a waveform model is suitable for GW data analysis. At higher mass ratios, regardless of the eccentricity, higher-order modes become increasingly important, and this is precisely where \model{} offers a significant improvement.

\subsection{Waveform comparison}
\label{sec:waveform_comparison}
We now present a series of waveform comparisons between the \model{} predictions and the corresponding NR data to visually demonstrate the modeling efficacy. In Figure~\ref{fig:case_0105}, we show the eccentric spherical harmonic modes obtained from the \model{} and the corresponding NR data for the case exhibiting the largest $L_2$-norm error among most of the higher-order modes. This simulation corresponds to a binary with mass ratio $q = 4$ and reference eccentricity $e_{\rm ref} = 0.282$. For comparison, we also display the higher-order modes obtained by applying the \texttt{gwNRHME} framework to the quasi-circular \texttt{NRHybSur3dq8} model, using the universal modulation extracted from the $(2,2)$ NR mode. We then show the results for another case exhibiting the largest $L_2$-norm error among higher-order modes in Figure~\ref{fig:case_0115}. In both cases, the \model{} predictions show  agreement with the NR results. In the first case, a small amount of dephasing is noticeable near the merger, but the overall agreement remains remarkably good.

\begin{figure}
\includegraphics[width=\columnwidth]{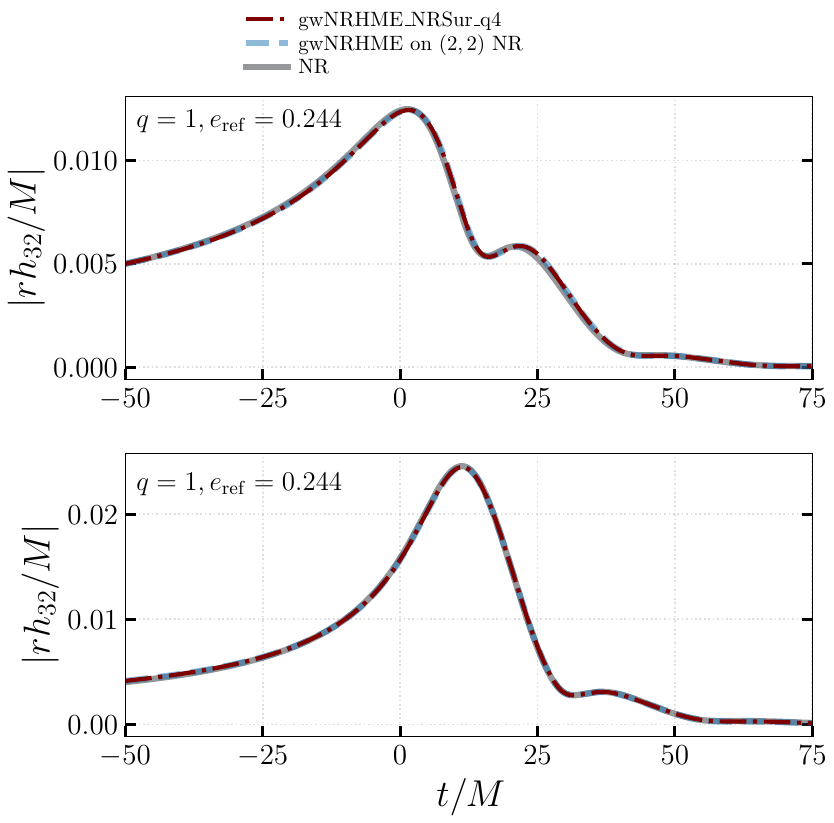}
\caption{We show the $(3,2)$-mode amplitude for two different binaries, comparing the \model{} predictions (maroon) with the corresponding NR data (grey) to demonstrate that \model{} can accurately capture mode mixing in the higher-order spherical harmonic modes. For reference, we also show the higher-order modes obtained by applying the \texttt{gwNRHME} framework to the quasi-circular \texttt{NRHybSur3dq8} model, using the universal modulation extracted from the $(2,2)$ NR mode (blue). More details are in Section~\ref{sec:results}.}
\label{fig:mode_mixing_32}
\end{figure}

We also find that \model{}, constructed using the \texttt{gwNRHME} framework, successfully reproduces intricate waveform features such as \textit{mode mixing} during the ringdown phase~\cite{Kelly:2012nd}, wherein a significant fraction of the dominant spherical harmonic mode’s power leaks into subdominant modes due to interference between spheroidal quasinormal modes with different damping rates. This effect is most prominent in the $(3,2)$ mode, where appreciable power from the dominant $(2,2)$ mode contaminates its signal. In Figure~\ref{fig:mode_mixing_32}, we illustrate this behavior for the $(3,2)$ mode in two binaries with different mass ratios and eccentricities. As evident from the NR data, the $(3,2)$-mode amplitude exhibits a characteristic oscillatory decay during ringdown, typical of mode mixing. 

Because the eccentricity is not large prior to plunge and the ringdown structure is not significantly altered by eccentricity for the simulations considered in this paper, waveforms obtained from \model{} can capture these features without any additional modifications, consistent with the behavior observed in the equal-mass eccentric surrogate model \texttt{NRSur2dq1Ecc} (see Fig.~9 of Ref.~\cite{Islam:2021mha}).

\begin{figure}
\includegraphics[width=\columnwidth]{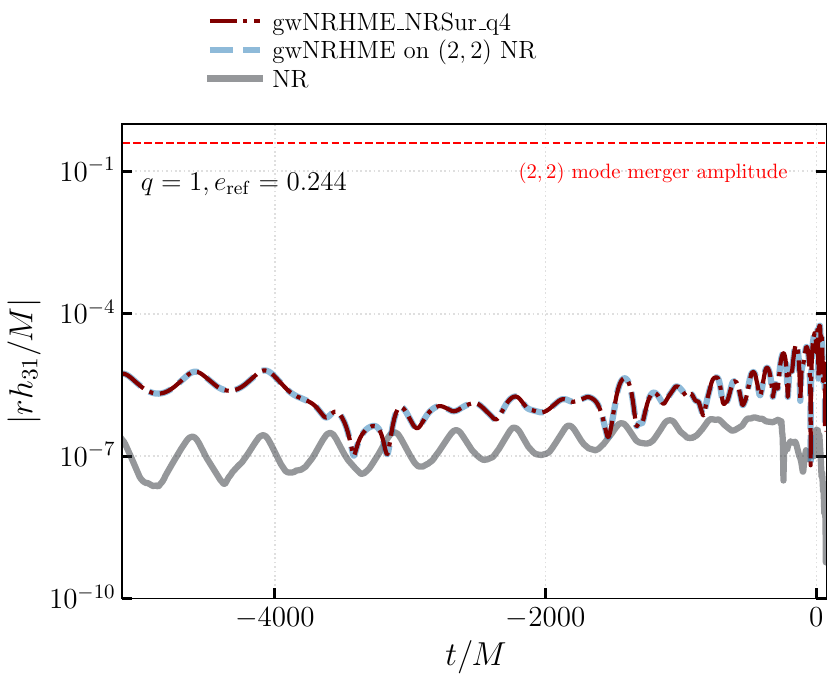}
\caption{We show the amplitude of a representative odd-$m$ mode, $(3,3)$, for an equal-mass BBH with reference eccentricity $e_{\rm ref} = 0.244$. Both the NR data (grey) and the \model{} predictions (maroon) remain consistent with zero within numerical accuracy, as indicated relative to the maximum $(2,2)$-mode amplitude (red dashed horizontal line). For comparison, we also show the amplitude obtained by applying the \texttt{gwNRHME} framework to the quasi-circular \texttt{NRHybSur3dq8} model, using the universal modulation extracted from the $(2,2)$ NR mode (blue). More details are in Section~\ref{sec:results}.}
\label{fig:mode_31_equalmass}
\end{figure}

\begin{figure}
\includegraphics[width=\columnwidth]{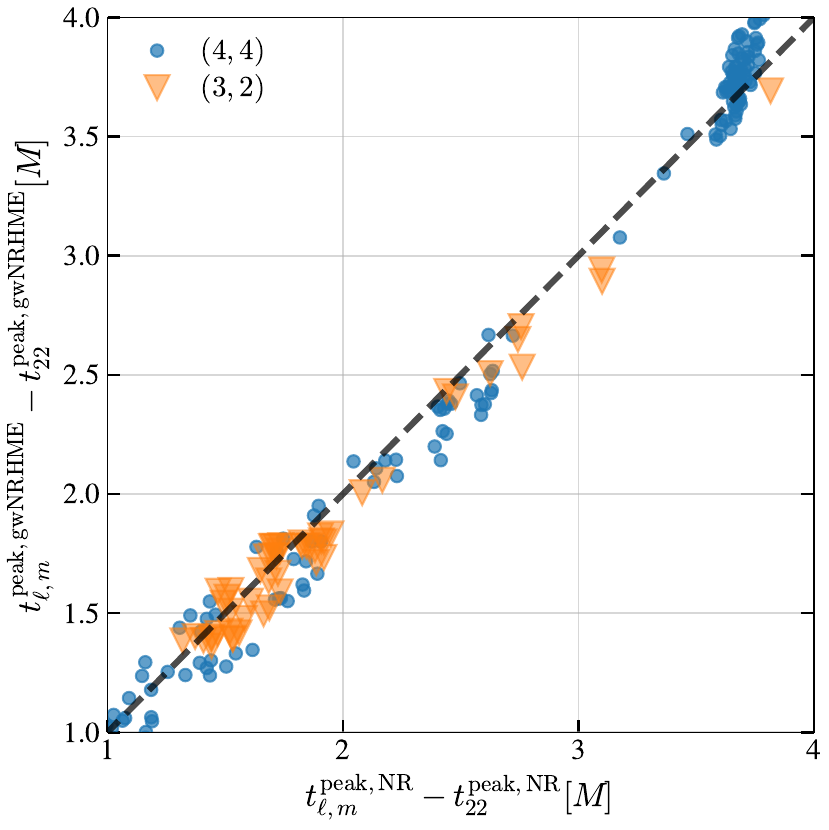}
\caption{We show that the relative peak times of two representative higher-order spherical harmonic modes, $(3,2)$ (orange triangles) and $(4,4)$ (blue circles), with respect to the $(2,2)$ mode, are consistent between the \model{} predictions and the NR data. The black dashed line indicates the line of equality for visual reference.}
\label{fig:gwnrhme_mode_peaks}
\end{figure}

Another important physical feature of higher-order waveform modes is their behavior in the equal-mass limit. Owing to symmetry, all odd-$m$ modes must vanish identically. However, this behavior is not explicitly enforced within the \texttt{gwNRHME} framework. The only information about this symmetry arises implicitly from the quasi-circular multi-modal waveform, which also exhibits zero radiation in these modes. We therefore perform a detailed consistency check to verify whether \model{} predictions reproduce this expected behavior and find that the amplitudes of these modes are indeed numerically consistent with zero. We demonstrate this in Figure~\ref{fig:mode_31_equalmass} for one representative case: the $(3,3)$ mode of an equal-mass binary. This behavior is similarly observed across all other odd-$m$ spherical harmonic modes. Note that while we test here how \texttt{gwNRHME} captures the identically zero odd-$m$ modes in the equal-mass limit, in the final model one can always explicitly enforce these modes to be zero by construction.

Finally, we examine the peak times of various spherical harmonic modes relative to the dominant $(2,2)$ mode. In Figure~\ref{fig:gwnrhme_mode_peaks}, we show that the relative peak times of two representative higher-order modes, $(3,2)$ (orange triangles) and $(4,4)$ (blue circles), are in agreement between the \model{} predictions and the NR data.

\begin{figure}
\includegraphics[width=\columnwidth]{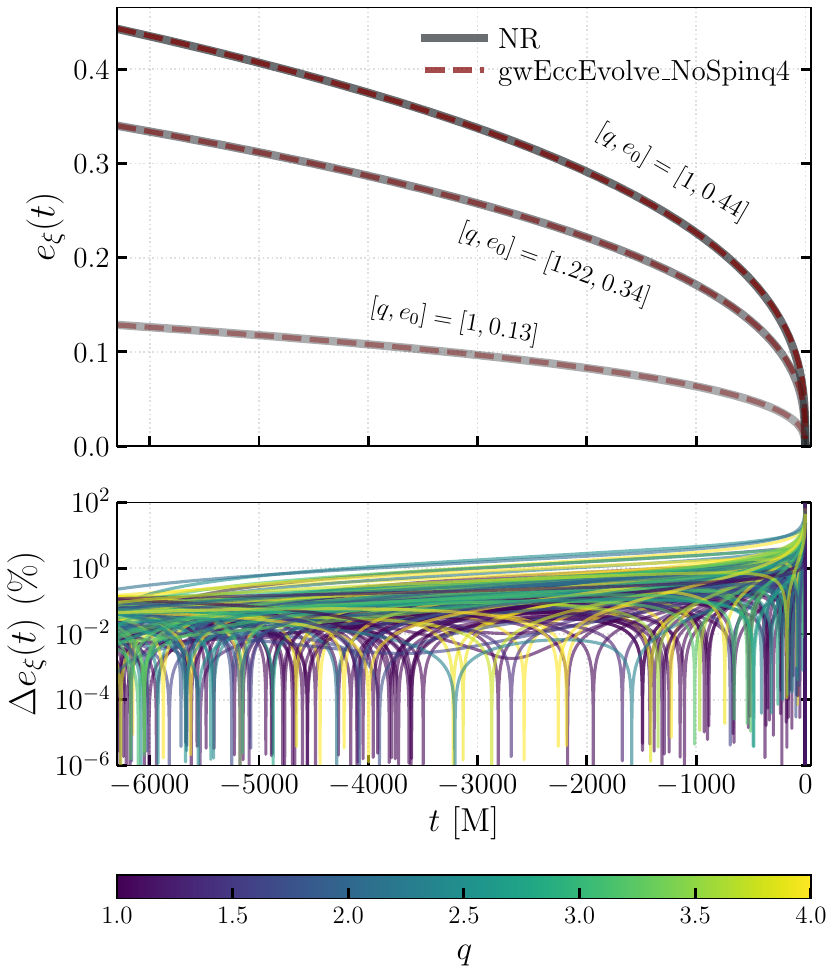}
\caption{(\textit{\textbf{Upper panel:}}) We show the NR eccentricity evolution $e_{\xi}(t)$ (solid grey lines; computed using the \texttt{gwModels} package, available at \href{https://github.com/tousifislam/gwModels}{https://github.com/tousifislam/gwModels}) for three representative simulations, alongside the corresponding validation predictions from the surrogate \texttt{gwEccEvolve\_NoSpinq4\_Sur} (dashed maroon lines) and analytical \texttt{gwEccEvv2} (dotted black lines) models. (\textit{\textbf{Lower panel:}}) Percentage error in $e_{\xi}(t)$ between the NR results and the \texttt{gwEccEvolve\_NoSpinq4} validation predictions (colorcoded by their mass ratio values). More details are in Section~\ref{sec:eccevolve_models}. 
}
\label{fig:ecc_xi_model_examples}
\end{figure}

\section{Model for eccentricity evolution}
\label{sec:eccevolve_models}
We now update the non-spinning eccentricity evolution model presented in Ref.~\cite{Islam:2025oiv} (\texttt{gwEccEvNS}) by incorporating the larger and more accurate set of NR simulations used in this work. As a first step, we project all eccentricity evolution time series $e_{\xi}(t)$ onto a common time grid, $t \in [-6200, 0]M$. We note that this time grid is slightly different from the one used in \texttt{NRSurE\_q4NoSpin\_22}; this is acceptable since, at present, there is no direct coupling between the two models. The primary goal of the eccentricity evolution model is to provide the community with a standalone tool to evolve eccentricity without the need for PN calculations or full waveform generation.
Once all eccentricity evolution time-series are defined on the same time grid, we construct a surrogate model for the scaled eccentricity evolution, $e_{\xi}(t)/e_{\xi,0}$, where $e_{\xi,0}$ is the initial eccentricity. We refer to this updated model as \texttt{gwEccEvolve\_q4NoSpin\_Sur}. 

To construct the surrogate model, we first compress the eccentricity evolution data using singular value decomposition (SVD) and retain the first two basis vectors to form a reduced-dimensional subspace. The projection coefficients corresponding to these basis vectors are then modeled as functions of the mass ratio $q$ and the initial eccentricity $e_{\xi,0}$ using GPR. To evaluate the performance of the surrogate, we carry out a 5-fold cross-validation analysis. To carry out the fitting and cross-validation procedures, we utilize the \texttt{scikit-learn} machine learning library~\cite{scikit-learn}.
We find that the surrogate accurately reproduces the eccentricity evolution across the parameter space (see Figure~\ref{fig:ecc_xi_model_examples}, upper panel), with fractional errors bounded within $5\%$ (for $t<=-100M$) for $95\%$ of the BBHs considered (Figure~\ref{fig:ecc_xi_model_examples}, lower panel). After $t = -100M$, the eccentricity becomes very small for most simulations, which often leads to larger relative percentage errors.

While \texttt{gwEccEvolve\_q4Spin\_Sur} provides an accurate description of the eccentricity evolution within its training domain, an alternative formulation is needed to handle cases with mass ratios and initial eccentricities outside the training region, as well as scenarios requiring a time grid longer than that used in training.  
To address this, we adopt the analytical ansatz introduced in \texttt{gwEccEvNS}~\cite{Islam:2025oiv} and refit it using the full set of NR simulations employed in this work.  
The resulting updated model, denoted as \texttt{gwEccEvNSv2}, is expressed as a function of the mass ratio $q$ and initial eccentricity $e_0$ as follows:
\begin{equation}
e_{\rm gwEccEvNSv2}(\tau, \tau_0, q, e_{0}) = e_{0} \times \left(\frac{\tau}{\tau_0}\right)^{n(q,e_0) / 48},
\label{eq:gwEccEvNS}
\end{equation}
where
\begin{align}
&n(q, e_0) =  n_1(q)\, n_2(e_0), \nonumber\\
&n_1(q) = -0.25027\, q + 18.31211, \nonumber\\
&n_2(e_0) = 1 + 0.38397\, e_0 - 1.93699\, e_0^2 + 2.0726\, e_0^3.
\end{align}
This updated model retains the analytical simplicity of the original formulation while incorporating the broader NR dataset to improve its accuracy across a wider range of parameter space. We find that the updated analytical model, \texttt{gwEccEvNSv2}, reproduces the eccentricity evolution with an accuracy comparable to the original \texttt{gwEccEvNS} model, achieving errors of approximately $\sim 5\%$ (for $t<=-100M$) for $95\%$ of the BBHs across the parameter space explored. Note that our analytical fit does not include contributions from the mean anomaly parameter. For very high eccentricities (not explored in this work), the initial mean anomaly may also influence the fit behavior. We leave this for future investigation.

\section{Adding eccentricity in other multi-modal quasi-circular models}
\label{sec:other_models}
While the results presented in Section~\ref{sec:results} employ \texttt{NRHybSur3dq8} as the base quasi-circular waveform model, it can be readily replaced with \texttt{NRSur7dq4}~\cite{varma2019surrogate} (in its non-spinning limit), another state-of-the-art NR surrogate model. This substitution is expected to yield comparable accuracy, as both \texttt{NRHybSur3dq8} and \texttt{NRSur7dq4} exhibit similar levels of agreement with NR data. 
Note that Ref.~\cite{Islam:2024zqo} demonstrated that \texttt{gwNRHME} yields highly accurate eccentric multi-modal waveform models only when both the underlying quasi-circular multi-modal model and the quadrupolar eccentric model are individually accurate. Any inaccuracy in either component, for instance, a less precise quadrupolar eccentric mode or less accurate higher-order modes in the quasi-circular model, can propagate into the combined waveform, reducing the overall accuracy. Since different NR surrogate quasi-circular base models are expected to produce comparable accuracy in the final eccentric model, we now turn our attention to models constructed within alternative modeling frameworks. In particular, we consider aligned-spin quasi-circular models developed within the EOB framework.

\begin{figure}
\includegraphics[width=\columnwidth]{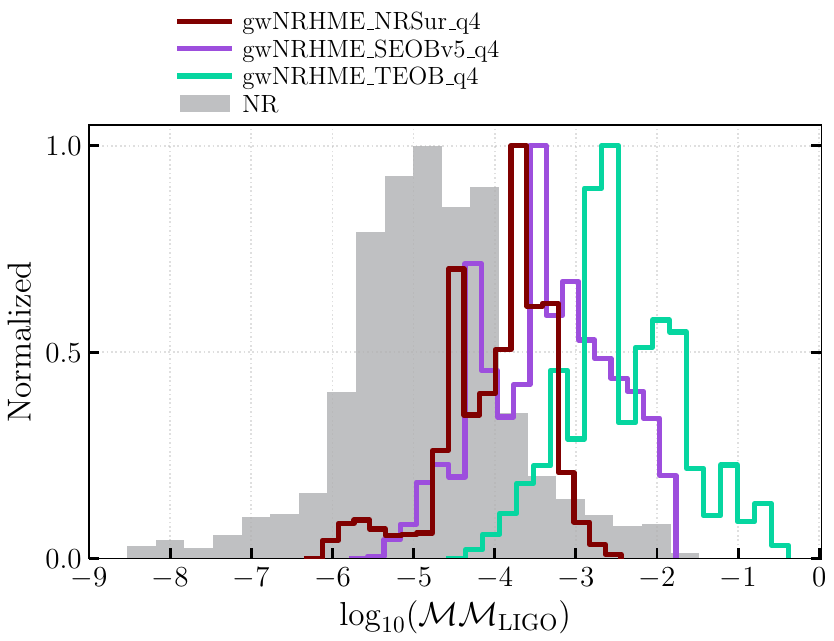}
\caption{We show frequency-domain Advanced LIGO mismatches between predictions from three different models constructed using the \texttt{gwNRHME} framework and the corresponding nonspinning, eccentric NR data. The mismatches are computed over different values of the orbital phase and inclination, as well as five detector-frame masses between $50\,M_{\odot}$ and $300\,M_{\odot}$. The maroon histogram shows mismatches for the \model{}, while the purple and teal histograms correspond to \texttt{gwNRHME\_SEOB\_q4} and \texttt{gwNRHME\_TEOB\_q4} models, respectively. All models include the same number of modes. For comparison, we also present mismatches obtained from NR simulations at two different numerical resolutions (grey histogram). More details are in Sec.~\ref{sec:other_models}.}
\label{fig:mismatches_seob_gwnrhme}
\end{figure}

We employ the \texttt{gwNRHME} framework to combine \texttt{NRSurE\_q4NoSpin\_22} with the aligned-spin EOB models \texttt{SEOBNRv5HM} and \texttt{TEOBResumS-Dali}, yielding a new eccentric waveform models denoted as \texttt{gwNRHME\_SEOBv5\_q4} and \texttt{gwNRHME\_TEOB\_q4} respectively.
Schematically, we write it as:
\begin{align}
\texttt{SEOBNRv5HM} 
&+ \texttt{NRSurE\_q4NoSpin\_22} 
+ \texttt{gwNRHME} \nonumber \\
&\longrightarrow \texttt{gwNRHME\_SEOBv5\_q4}
\end{align}
and
\begin{align}
\texttt{TEOBResumS-Dali} 
&+ \texttt{NRSurE\_q4NoSpin\_22} 
+ \texttt{gwNRHME} \nonumber \\
&\longrightarrow \texttt{gwNRHME\_TEOB\_q4}.
\end{align}
We assess the accuracy of these models using the same procedures outlined in Section~\ref{sec:results}. 
In Figure~\ref{fig:mismatches_seob_gwnrhme}, we present the frequency-domain mismatches between the \texttt{gwNRHME\_SEOBv5\_q4} model predictions and NR data, computed by including six spherical harmonic modes and assuming the Advanced LIGO design sensitivity curve. The resulting mismatches are predominantly in the range of $10^{-5}$ to $10^{-2}$, which is roughly an order of magnitude higher than those obtained with \model{}. This difference arises because, although \texttt{SEOBNRv5HM} is an accurate model for quasi-circular aligned-spin binaries and widely regarded as state-of-the-art within the EOB framework, its accuracy is still somewhat lower than that of \texttt{NRHybSur3dq8}~\cite{Pompili:2023tna}. Nevertheless, these results further confirm the robustness and modularity of the \texttt{gwNRHME} framework in constructing eccentric waveform models from different quasi-circular baselines. 
For \texttt{gwNRHME\_TEOB\_q4}, we find that the mismatches are almost an order of magnitude larger than the \texttt{gwNRHME\_SEOBv5\_q4} and \texttt{gwNRHME\_NRSur\_q4} models.
 Overall, \texttt{gwNRHME\_SEOBv5\_q4} and \texttt{gwNRHME\_TEOB\_q4} yield median mismatches of $\sim 2\times10^{-4}$ and $\sim 10^{-3}$, respectively, with standard deviation of $\sim 2 \times 10^{-3}$ and $\sim 2 \times 10^{-2}$ respectively. These values are larger than \model{} which achieves median mismatches of $\sim 9\times 10^{-5}$, with a standard deviation of $\sim 2 \times 10^{-4}$.
 
This demonstrates the simplicity and modularity of the \texttt{gwNRHME} framework in extending existing quasi-circular multi-modal models to their eccentric counterparts using \texttt{NRSurE\_q4NoSpin\_22} as the eccentric quadrupolar base. We expect that \texttt{NRSurE\_q4NoSpin\_22} can similarly be coupled with other quasi-circular models such as \texttt{IMRPhenomD}~\cite{Khan:2015jqa,Husa:2015iqa} and \texttt{IMRPhenomTHM}~\cite{Estelles:2020twz} following the same strategy employed in constructing the \texttt{NRTidal} extensions~\cite{Dietrich:2017aum,Dietrich:2019kaq} where tidal corrections are added to any BBH waveform models in a modular fashion.

\begin{figure}
\includegraphics[width=\columnwidth]{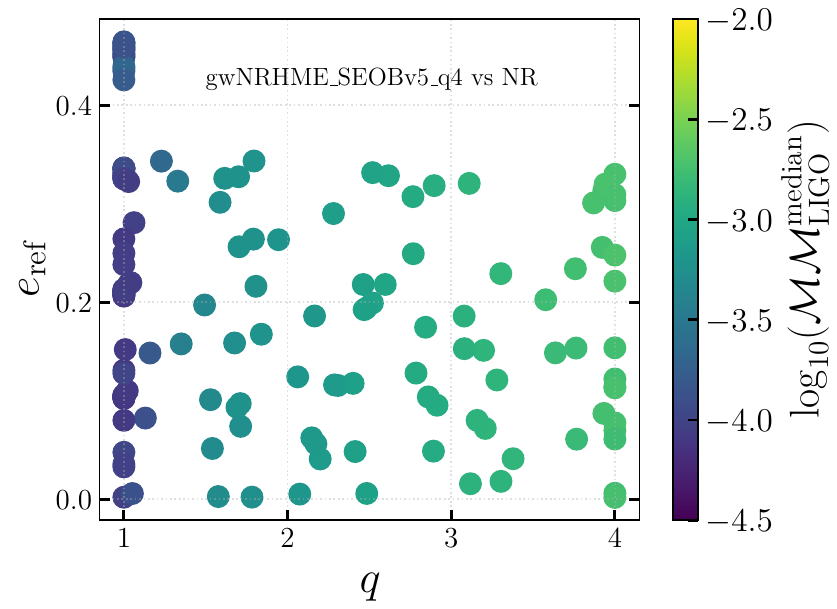}
\includegraphics[width=\columnwidth]{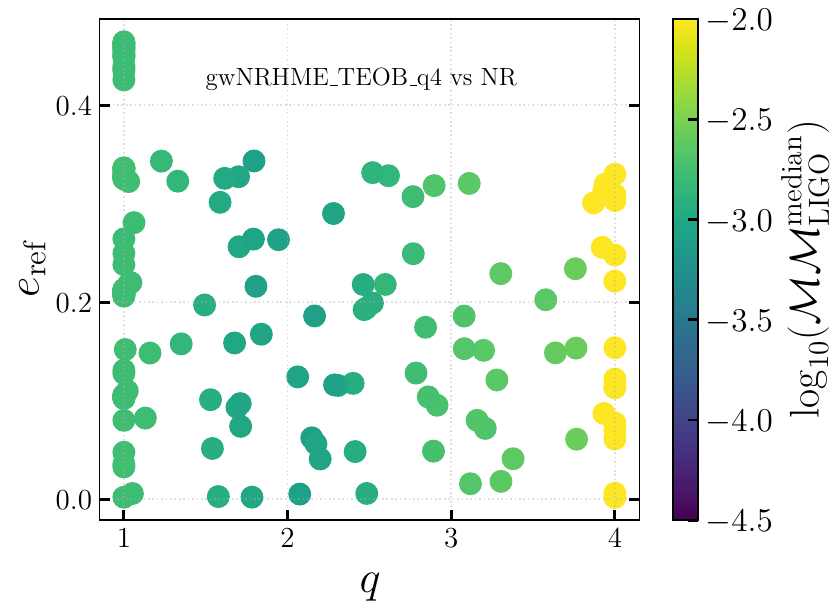}
\caption{We show median Advanced LIGO mismatches between \texttt{gwNRHME\_SEOBv5\_q4} (\texttt{gwNRHME\_TEOB\_q4}) predictions and NR data as a function of the mass ratio and the reference eccentricity in the upper (lower) panel. The median is computed over different values of the orbital phase and inclination. More details are in Sec.~\ref{sec:other_models}.}
\label{fig:mismatches_eob_paramspace_colorbar}
\end{figure}

While \texttt{gwNRHME\_SEOBv5\_q4} and \texttt{gwNRHME\_TEOB\_q4} achieves reasonably good overall mismatches, it is also good to understand where the model performs relatively poorly.
In Fig.~\ref{fig:mismatches_eob_paramspace_colorbar}, we therefore show the median Advanced LIGO mismatches between the predictions of \texttt{gwNRHME\_SEOB\_q4} (\texttt{gwNRHME\_TEOB\_q4}) and the NR waveforms as a function of mass ratio and reference eccentricity. We find that both \texttt{gwNRHME\_SEOB\_q4} and \texttt{gwNRHME\_TEOB\_q4} exhibit a loss of accuracy with increasing mass ratio, consistent with the behavior observed for \model{}.

\section{Summary of models}
\label{sec:summary}
In summary, we provide the following models for eccentric binary black holes mergers.
\begin{itemize}
    \item \model{}: A model for eccentric, non-spinning binary black hole waveforms valid over the parameter ranges $q \in [1,4]$ and $e_{\rm ref} \in [0.001, 0.43]$. This model is constructed combining \texttt{NRSurE\_q4NoSpin\_22} with \texttt{NRHybSur3dq8}. The model includes a total of nine spherical harmonic modes: $(\ell, m) = \{(2,2), (2,1), (3,1), (3,2), (3,3), (4,2), (4,3), (4,4), (5,5)\}$ (cf. Sections~\ref{sec:results}).
    \item \texttt{gwNRHME\_SEOBv5\_q4}: A model for eccentric, non-spinning binary black hole waveforms valid over the parameter ranges $q \in [1,4]$ and $e_{\rm ref} \in [0.001, 0.43]$. This model is constructed combining \texttt{NRSurE\_q4NoSpin\_22} with \texttt{SEOBNRv5HM}. The model includes a total of six spherical harmonic modes: $(\ell, m) = \{(2,2), (2,1), (3,2), (3,3), (4,3), (4,4)\}$. This model can be straightforwardly extrapolated to systems with moderate spins (cf. Section~\ref{sec:other_models}).
    \item \texttt{gwNRHME\_TEOB\_q4}: A model for eccentric, non-spinning binary black hole waveforms valid over the parameter ranges $q \in [1,4]$ and $e_{\rm ref} \in [0.001, 0.25]$. This model is constructed combining \texttt{NRSurE\_q4NoSpin\_22} with \texttt{TEOBResumS-Dali}. The model includes a total of six spherical harmonic modes: $(\ell, m) = \{(2,2), (2,1), (3,1), (3,2), (3,3), (4,2), (4,3), (4,4), (5,5)\}$. This model can be straightforwardly extrapolated to systems with moderate spins (cf. Section~\ref{sec:eccevolve_models}).
    \item \texttt{gwEccEvolve\_NoSpinq4\_Sur}: A surrogate model for the eccentricity evolution, $e_{\xi}(t)$, for non-spinning eccentric binary black holes. The model is valid for mass ratios $q \in [1,4]$ and initial eccentricities $e_{0} \in [0.001, 0.43]$, and describes the last $6000M$ of the binary evolution up to merger.
    \item \texttt{gwEccEvNSv2}: A phenomenological model for the eccentricity evolution, $e_{\xi}(t)$, for non-spinning eccentric binary black holes. The model is designed to accept arbitrary mass ratios, initial eccentricities, and time grids. However, the model parameters are calibrated using the same NR dataset employed for \texttt{gwEccEvolve\_NoSpinq4} (cf. Section~\ref{sec:eccevolve_models})~\footnote{Model performance at large mass ratios and high eccentricities is not well tested at this point.}.
\end{itemize}

\section{Concluding remarks}
\label{sec:conclusion}
We presented \model{}, a multi-modal, non-spinning eccentric waveform model valid for mass ratios $q \in [1,4]$ and eccentricities up to $0.43$ when measured with Eq.~\eqref{eq:exi}. The model is constructed by combining the multi-modal quasi-circular surrogate \texttt{NRHybSur3dq8} with the eccentric quadrupolar surrogate \texttt{NRSurE\_q4NoSpin\_22}. In its current form, \model{} includes the following spherical harmonic modes: $(\ell, m) = \{(2,2), (2,1), (3,1), (3,2), (3,3), (4,2), (4,3), (4,4), (5,5)\}$. Additional modes can be incorporated straightforwardly by applying the universal eccentric modulation functions to the corresponding quasi-circular modes.
We find that \model{} reproduces NR waveforms with high fidelity, yielding median Advanced LIGO frequency-domain mismatches typically approximately $9 \times 10^{-5}$. 

Owing to the modular nature of the \texttt{gwNRHME} framework, eccentricity can be projected onto any quasi-circular model with minimal effort. As a demonstration, we combine \texttt{NRSurE\_q4NoSpin\_22} with two EOB models (\texttt{SEOBNRv5HM} and \texttt{TEOBResumS-Dali}, in their non-spinning limits), resulting in NR-faithful eccentric waveform models with median mismatches of $2\times10^{-4}$, and $10^{-3}$, respectively.
The \texttt{gwNRHME} framework is publicly available through the \texttt{gwModels} package, and the resulting waveform models will be released via the \texttt{gwsurrogate} package.

There are several directions for future work. One immediate next step is to construct fully data-driven eccentric surrogate models for the higher-order modes, following the methodology outlined in Refs.~\cite{Nee:2025nmh,Maurya:2025shc}. In the near term, we anticipate that \model{} can be employed in GW data analysis. Moreover, we envision the \texttt{gwNRHME} framework providing accurate eccentric extensions to a variety of existing quasi-circular models by leveraging the modulation functions extracted from the \texttt{NRSurE\_q4NoSpin\_22} model.

Looking ahead, we plan to extend our surrogate modeling framework to aligned-spin and precessing eccentric binaries. At least for aligned-spin systems, we expect the \texttt{gwNR(X)HME} framework to achieve a comparable level of accuracy as demonstrated in the non-spinning case.

We also note that the empirical universality of the eccentric modulation functions extracted from different modes~\cite{Islam:2024rhm,Islam:2024bza} remains primarily phenomenological at this stage. Although this universality has been validated for mass ratios up to $q=5$ and eccentricities up to $\sim 0.5$~\cite{Islam:2024rhm,Islam:2024bza} , additional tests are required to confirm its robustness across a broader range of BBH configurations. It is conceivable that for systems with very high mass ratios or large eccentricities, extensions may be necessary beyond the current \texttt{gwNR(X)HME} framework.
We leave a detailed investigation of this extension to future work.

\begin{acknowledgments}
We are grateful to Tejaswi Venumadhav, Jay Wadekar and Carl-Johan Haster for valuable discussions. 
We thank Lorenzo Pompili, Carlos Lousto and Khun Sang Phukon for helpful comments on the manuscript.
This research was supported in part by the National Science Foundation under Grant No. NSF PHY-2309135 and the Gordon and Betty Moore Foundation Grant No. GBMF7392.
Use was made of computational facilities purchased with funds from the National Science Foundation (CNS-1725797) and administered by the Center for Scientific Computing (CSC). The CSC is supported by the California NanoSystems Institute and the Materials Research Science and Engineering Center (MRSEC; NSF DMR 2308708) at UC Santa Barbara. 
S.F. acknowledges support from NSF Grants No. AST-2407454 and PHY-2110496.
This material is based upon work supported by the National Science Foundation under Grants No. PHY-2407742, No. PHY-2308615, and No. OAC-2513338, and by the Sherman Fairchild Foundation at Cornell.
This work was partly supported by UMass Dartmouth's Marine and Undersea Technology (MUST) research program funded by the Office of Naval Research (ONR) under grant no. N00014-23-1-2141. Some computations were performed on the UMass-URI UNITY HPC/AI cluster at the Massachusetts Green High-Performance Computing Center (MGHPCC).
\end{acknowledgments}

\bibliography{References}

\end{document}